\documentclass[11pt]{article}
\usepackage[T1]{fontenc}
\usepackage{lmodern}
\usepackage[a4paper,margin=1in]{geometry}
\usepackage{microtype}
\usepackage{amsmath,amssymb}
\usepackage{graphicx}
\usepackage{booktabs}
\usepackage{tabularx}
\usepackage{array}
\usepackage{multirow}
\usepackage{makecell}
\usepackage{threeparttable}
\usepackage{longtable}
\usepackage{enumitem}
\usepackage{xcolor}

\usepackage{wasysym}
\usepackage{caption}
\usepackage[numbers,sort&compress]{natbib}
\usepackage[hidelinks]{hyperref}
\usepackage{url}
\makeatletter\g@addto@macro\UrlBreaks{\do\-\do\_}\makeatother
\setlist{itemsep=2pt,topsep=4pt}
\newcommand{\sysF}{System~F}
\newcommand{\ev}[1]{\textsuperscript{[#1]}}
\newcommand{\full}{\CIRCLE}
\newcommand{\half}{\LEFTcircle}
\newcommand{\none}{\Circle}
\newcolumntype{L}[1]{>{\raggedright\arraybackslash}p{#1}}
\newcolumntype{Y}{>{\raggedright\arraybackslash}X}

\title{Beyond the Agent Harness: Boundary Criteria for Software-Factory Orchestration\\[4pt]
\large A Single-Operator Case Study}
\author{Lei Hua, PhD\\{}Independent Researcher\\{}\texttt{lhua0420@gmail.com}}
\date{November 2026}

\begin{document}
\maketitle

\begin{abstract}
Several recent systems add a layer above the coding-agent harness that organises many agent runs into continuous delivery. The existence of such a layer is not our contribution. What is missing is a rule for deciding which layer a mechanism or a failure belongs to. We propose boundary criteria based on an operational definition of a run. A mechanism confined to one run belongs to the agent harness. One that coordinates runs, hands state from one run to a later one, or is updated from run outcomes belongs to a \emph{factory harness}. One that must change with the application domain belongs to the domain layer. Two residual tests assign infrastructure and model causes. For failures, we code where the defect originated, which control failed to stop it, and where it was fixed as three separate fields, and we require positive evidence before attributing a failure to the environment. We developed the criteria on documented failures from a single operator's pipeline serving six projects and on 26 components of three published systems. We then froze them and had two model coders of the same family apply them to the same descriptions of 46 held-out units. They agreed on all 26 mechanism layers and on 18--19 of 20 incident fields ($\kappa$ 0.86--0.93), but both coders recorded the incident as undetermined or unrecorded in 8--13 of the 20 cases per field, and each flagged half of all units as ambiguous, mostly for missing information. The check exposed conflicts in the rules; a revision that separates mechanism classification from incident attribution resolves them but, in a sensitivity re-coding, agreed less ($\kappa$ 0.78--0.93). Agreement here is neither accuracy nor usefulness, and human coders have not yet been tested. Case evidence and seven hypotheses are reported as preliminary.
\end{abstract}

\section{Introduction}
\label{sec:intro}

A common formulation in the practitioner literature holds that an agent consists of a model and a harness, where the harness is everything that is not the model: tools, context management, permissions, sandboxing and the loop that drives the model~\citep{langchain2026anatomy,nandyala2026engineering}. Harness engineering has since become a topic of its own, with practitioner frameworks~\citep{bockeler2026harness,thoughtworks2026radar}, a source-code study of eleven production coding harnesses~\citep{barbaste2026harness}, a survey and taxonomy~\citep{li2026harnesssurvey} and a constitutive definition with an inclusion--exclusion test~\citep{demacedo2026harness}.

In parallel, organisations describe \emph{software factories} in which many agent runs are organised into continuous delivery~\citep{shapiro2026fivelevels,strongdm2026factory,medisetty2026uber,factory2026factory2,pineyro2026pento}. Several authors already speak of the harness ``becoming'' a factory~\citep{bhanot2026harness4,worthingtonbodart2026harness}, and recent systems implement an explicit layer above existing harnesses: Harness-of-Harness (HoH) organises a fixed harness--model configuration into planning--coding--testing loops with cross-loop state~\citep{yan2026hoh}; Sengupta et al.\ describe a contract-driven meta-engineering harness with a failure arbiter and outer-loop calibration~\citep{sengupta2026metaeng}; Omnigent composes and controls several harnesses as a meta-harness~\citep{databricks2026omnigent}; Proof-or-Stop gates lifecycle transitions on hash-bound evidence~\citep{huang2026proofstop}.

The existence of a layer above the agent harness is therefore established, and we do not claim it. What remains unclear is where the boundary lies. The survey by Li et al.\ places cross-run persistence, multi-agent orchestration and issue-to-PR pipelines \emph{inside} the harness~\citep{li2026harnesssurvey}; de Macedo's test classifies an orchestrator or an evaluation harness as neighbours \emph{outside} it~\citep{demacedo2026harness}; Sengupta et al.\ call the whole production system a harness and reserve ``factory'' for the larger system including its accumulated assets~\citep{sengupta2026metaeng}. These are different usages of the same words, and none of them gives a rule for assigning a given mechanism (a resource lock, a protected test file, a retry budget, a preflight check) to a layer. We expect this to matter for design, because mechanisms in different layers plausibly change at different rates and for different reasons (H4, H5), and for evaluation, because a comparison of agent harnesses that ignores the outer layer can attribute to the harness effects that come from verification or scheduling.

This paper makes the following contributions, of which the first two are the main ones.
\begin{enumerate}[label=(\roman*)]
  \item \textbf{Boundary criteria and a three-field incident coding.} Criteria that assign a mechanism to the agent harness, the factory harness or the domain layer, residual tests for plant and model, an operational definition of a run, and a coding of failures that separates defect origin, control gap and fix location, with an evidence rule for environmental causes (Section~\ref{sec:criteria}, Appendix~\ref{app:codebook}). The criteria operationalise an informal distinction already drawn by practitioners~\citep{pineyro2026pento}.
  \item \textbf{A development-then-validation record for the criteria.} The criteria were developed on documented failures from our case and on the components of three published systems, with the ambiguous cases reported. They were then frozen as version~1.0 and applied to 46 held-out units by two independent model coders. The conflicts this exposed led to a revised version~1.1, which separates mechanism classification from incident attribution; a re-coding under v1.1 is reported as a sensitivity analysis, not a validation (Section~\ref{sec:apply}).
  \item \textbf{A five-layer stack} that separates the domain layer (above) and the plant layer (below) from the factory harness (Section~\ref{sec:stack}).
  \item \textbf{A single-operator, multi-project case} (Sections~\ref{sec:case} and~\ref{sec:evidence}). A factory-layer pipeline, \sysF, serves six projects with three selectable agent harnesses, shared runners and resource locks. To our knowledge, none of the 14 works compared in Table~\ref{tab:related-work} (arXiv, OpenReview, vendor and practitioner blogs, searched up to 2026-11-02) describes this setting. The case also supplies preliminary evidence: a 48-run same-condition backend comparison recomputed from per-run data, held-out-check comparisons on two real defects, an episode in which a self-test result changed with the sandbox, and a rejection matrix for verification receipts signed outside the agent's sandbox. The receipt mechanism is close to Proof-or-Stop~\citep{huang2026proofstop}.
  \item \textbf{An evaluation agenda} of seven hypotheses with decision rules that distinguish support, refutation and inconclusive results (Section~\ref{sec:hyp}).
\end{enumerate}

We follow case-study reporting guidelines for software engineering~\citep{runeson2009casestudy}. The system designer, operator and evaluator are the same person, the samples are small and adjudication was not blinded (Section~\ref{sec:threats}). Every case number is keyed to a de-identified data appendix (Appendix~\ref{app:data}); the E1 numbers were recomputed from per-run data, and the remaining case numbers are author-reported. \emph{Naming note:} ``Factory'' is also the name of a company (factory.com; formerly factory.ai) whose product narrative uses the term ``software factory''~\citep{factory2026factory2}. This work has no affiliation with that company; we use the term generically.

\section{Background and Related Work}
\label{sec:related}

\paragraph{Agent harnesses and harness engineering.}
The ReAct pattern, in which a model interleaves reasoning and actions in a loop~\citep{yao2023react}, underlies the coding agents analysed by Barbaste et al., who map eleven production harnesses onto seven subsystems (agent loop, LLM integration, tools, memory and context, safety and permissions, orchestration, extensibility) plus an interface layer and a session substrate~\citep{barbaste2026harness}. That study already covers sub-agent orchestration and cross-session memory and names an ``outer verification loop'' pattern, so harness-internal mechanisms are not confined to a single run in its usage; what it does not analyse is delivery-level concerns such as scheduling across projects or verification outside the agent's reach. SWE-agent shows that the design of the agent--computer interface affects performance on benchmarks such as SWE-bench~\citep{yang2024sweagent,jimenez2024swebench}; OpenHands provides an open platform for such agents with sandboxed execution~\citep{wang2025openhands}. B{\"o}ckeler separates feedforward \emph{guides} from feedback \emph{sensors} and computational from inferential controls~\citep{bockeler2026harness}; Thoughtworks applies the split to delivery pipelines~\citep{nandyala2026engineering,thoughtworks2026decoder} and recommends wiring deterministic quality gates into agent workflows~\citep{thoughtworks2026radar}. Note that these sensors are designed to feed back \emph{to} the agent; the verification we study in Section~\ref{sec:e4} is deliberately \emph{unreachable} by it.

\paragraph{Definitions and taxonomies.}
De Macedo proposes four constitutive conditions for an agent harness (a reasoning--action--observation loop, a tool interface that can alter the environment, task-aware context management, and at least one control mechanism independent of the model) and uses them to separate a harness from a framework, an SDK, an IDE plugin, an evaluation harness and an orchestrator~\citep{demacedo2026harness}. We adopt this inclusion--exclusion style of argument and move it one level up: de Macedo decides whether a system \emph{is} an agent harness; our criteria decide \emph{which layer} a mechanism belongs to. Li et al.\ propose a seven-layer taxonomy (Execution, Tooling, Context, Lifecycle, Observability, Verification, Governance) and map more than 170 open-source projects onto it; their taxonomy places cross-run persistence and multi-agent orchestration inside the harness~\citep{li2026harnesssurvey}. HarnessFix attributes agent failures to harness artefacts using that taxonomy and repairs them~\citep{chen2026harnessfix}. We return to the disagreement with Li et al.\ in Section~\ref{sec:disc}.

\paragraph{Layers above the harness.}
HoH runs planner, developer and QA-tester invocations of the same harness--model configuration, carries an artefact state and an evidence state across loops, freezes the candidate for independent QA, and reports gains over standalone harnesses on three benchmarks~\citep{yan2026hoh}. Sengupta et al.\ describe a seven-layer production architecture with two-pass contract compilation, persistent markdown memory, domain specialisation records, independence- and attention-based verification, a four-way failure arbiter (bug, spec gap, noise, contract ambiguity) and a calibration loop, evaluated as an early deployment of 17 features~\citep{sengupta2026metaeng}. Omnigent adds cross-harness composition, stateful policies and sandboxing above several harnesses~\citep{databricks2026omnigent}. Proof-or-Stop treats agent outputs as claims and admits a lifecycle transition only on fresh evidence bound to hashes of source state, policy and command set; its offline test rejects 18 tamper classes~\citep{huang2026proofstop}. Role-based multi-agent frameworks such as MetaGPT encode standard operating procedures as agent roles within one workflow~\citep{hong2024metagpt}; HoH and Sengupta et al.\ differ in carrying state and verification across runs. Practitioner essays describe the same shift: Pento distinguishes a harness that runs the agent from a factory that runs many agents against a backlog~\citep{pineyro2026pento}, an informal version of our criteria C1 and C2; Bhanot and Worthington-Bodart describe the harness turning into a factory~\citep{bhanot2026harness4,worthingtonbodart2026harness}.

\paragraph{Software factories.}
Shapiro's five levels end in a ``dark factory'' in which humans neither write nor review code~\citep{shapiro2026fivelevels}; StrongDM states the same two rules for its factory~\citep{strongdm2026factory}. Factory describes the software factory as a continuous loop from external signals through delivery and monitoring back to new signals~\citep{factory2026factory2}. Uber reports that more than 70\% of its pull requests are \emph{attributed} to agents (an attribution measure, not a share of code) and more than 30K agent skill executions per day~\citep{medisetty2026uber}. Anthropic describes rebuilding source control for agent-heavy workloads~\citep{anthropic2026scion}, and an OpenAI engineering post reports an agent-written codebase built by a small team~\citep{lopopolo2026harness}. These sources are vendor or practitioner reports; we cite them for positioning, not as evidence.

\paragraph{Verification, provenance and CI.}
Signed provenance and verifiable supply chains are established: in-toto verifies each supply-chain step cryptographically~\citep{torresarias2019intoto}; SLSA's build levels require that signing material be inaccessible to user-defined build steps~\citep{slsa2023spec}; Sigstore makes signing widely usable~\citep{newman2022sigstore}; reproducible builds let third parties check that binaries match sources~\citep{lamb2022reproducible}. Continuous-integration research documents CI's costs and benefits~\citep{hilton2016ci} and the prevalence of flaky tests whose outcome changes without code changes~\citep{luo2014flaky}, both relevant to separating environment failures from quality failures. Agents with access to tests can modify or delete them to pass~\citep{zhong2025impossiblebench}, and LLM judges show systematic biases~\citep{zheng2023judging}; both motivate verification the agent cannot reach and caution about LLM adjudication. Operational practice in site reliability engineering~\citep{beyer2016sre} informs the plant layer.

Table~\ref{tab:related-work} compares representative works on five properties.

\providecommand{\full}{\CIRCLE}
\providecommand{\half}{\LEFTcircle}
\providecommand{\none}{\Circle}
\begin{table}[!htbp]
\centering
\small
\caption{Representative related work compared on five properties. Cells record only what each source states (one justification per cell is given in the supplementary citation log). For the column \emph{single-agent / single-run scope}, \full{} means the work is scoped to one agent's run; for the other columns, \full{} means the property is stated explicitly.}
\label{tab:related-work}
\begin{tabularx}{\linewidth}{@{}>{\raggedright\arraybackslash}p{3.3cm} l c c c c c >{\raggedright\arraybackslash}X@{}}
\toprule
Work & Type & S & P & V & D & B & Basis (paraphrased) \\
\midrule
LangChain (Anatomy of an Agent Harness)~\cite{langchain2026anatomy} & Blog & \half & \half & \half & \none & \half & Defines harness as all non-model code; verification is a harness design choice \\
Barbaste et al. (source-code study)~\cite{barbaste2026harness} & Preprint & \half & \half & \half & \none & \half & 7 subsystems inside one harness runtime; Omnigent as contrast \\
Thoughtworks / B\"ockeler (guides \& sensors)~\cite{bockeler2026harness,nandyala2026engineering,thoughtworks2026radar} & Blog/Radar & \half & \none & \half & \half & \half & Guides (feedforward) vs.\ sensors (feedback) around a coding agent \\
de Macedo (harness definition)~\cite{demacedo2026harness} & Preprint & \full & \none & \none & \none & \half & Necessary and sufficient conditions; inclusion/exclusion test \\
Li et al.\ (harness survey, ETCLOVG)~\cite{li2026harnesssurvey} & Survey & \half & \half & \half & \none & \half & Seven-layer taxonomy; files cross-run persistence and pipelines inside the harness \\
Databricks Omnigent (meta-harness)~\cite{databricks2026omnigent,barbaste2026harness} & Blog/OSS & \none & \full & \half & \none & \half & Composition, control (stateful policies, budgets), shared sessions above harnesses \\
Harness-of-Harness (HoH)~\cite{yan2026hoh} & Preprint & \none & \full & \full & \half & \none & Planner/Developer/QA loops over existing harnesses; multi-day runs \\
Meta-engineering harness (Sengupta et al.)~\cite{sengupta2026metaeng} & Preprint & \none & \full & \half & \half & \none & Contracts, independent/adversarial verification, failure arbiter, calibration \\
Proof-or-Stop (Huang et al.)~\cite{huang2026proofstop} & Preprint & \none & \full & \half & \none & \half & Agent outputs as claims; source-state-bound receipts gate lifecycle transitions \\
Factory 2.0~\cite{factory2026factory2} & Vendor & \none & \half & \none & \none & \none & Signal-to-production feedback loop; model independence; sovereignty \\
StrongDM Software Factory$^{\ddagger}$~\cite{strongdm2026factory} & Vendor/spec & \none & \half & \half & \none & \none & No human-written or human-reviewed code; scenario holdouts; digital twins \\
Uber software factory~\cite{medisetty2026uber} & Industry blog & \none & \half & \none & \none & \none & >70\% of PRs agent-attributed; skills registry; cost equation \\
Pento AI~\cite{pineyro2026pento} & Practitioner & \none & \none & \half & \none & \half & Harness runs one agent; factory runs many against a backlog \\
Bhanot; Worthington-Bodart~\cite{bhanot2026harness4,worthingtonbodart2026harness} & Blogs & \none & \half & \half & \none & \none & 'Harness becomes a factory'; improve the system, not the output \\
\midrule
This paper (Factory Harness)$^{\dagger}$ & Case study & \none & \full & \half & \full & \full & Five-layer stack; three boundary criteria; single-operator evidence \\
\bottomrule
\end{tabularx}

\vspace{2pt}
\begin{minipage}{\linewidth}\footnotesize
S: scope limited to a single agent / single run. P: persistent state across runs. V: verification placed outside the agent's reach. D: domain knowledge as a separable, pluggable layer. B: explicit criteria for assigning a mechanism to a layer.
\full{} stated; \half{} partial, implicit, or stated for a different boundary; \none{} not addressed in the material we read.
$^{\dagger}$Row describes what this paper \emph{proposes} and implements in a single-operator system; it is not an external assessment, and generality (H4) is untested.
$^{\ddagger}$Verification cell based on secondary reports of the StrongDM site (primary page not read), hence \half{} rather than \full.
Bhanot~\cite{bhanot2026harness4} was read only in excerpts; its cells are provisional. The same rule is applied to Proof-or-Stop and to this paper: V is \half{} when verification is outside the agent's sandbox but the trust model is limited (Proof-or-Stop: local key, single host, compromised runner excluded; this paper: checks and publication share a VM and system user, and the scheduler holds an administrator token). Proof-or-Stop's receipt mechanism closely anticipates this paper's signed-receipt design. Pento's harness/factory distinction is an informal precursor of criteria C1/C2.
\end{minipage}
\end{table}

\section{The Factory Harness Framework}
\label{sec:framework}

\subsection{Definition}
\label{sec:def}
\begin{quote}
A \textbf{factory harness} is the layer between the agent harness and the domain layer that organises the runs of one or more agents into a production process that is \emph{verifiable, recoverable and continuously deliverable}. It is responsible for scheduling and resource locks, persistent state across runs, verification that the agent cannot reach, failure classification and repair budgets, environment preflight, and human intervention points. Its structure is intended to be domain-independent; its rules and evaluations evolve from observed failures, which we hypothesise come increasingly from the domain once the structure has converged (H4, H5).
\end{quote}
The definition is analytical, not lexical. In the usage of Li et al.~\citep{li2026harnesssurvey} the factory harness corresponds to the cross-run part of the harness; in that of Sengupta et al.~\citep{sengupta2026metaeng} it covers their orchestration, verification and calibration layers. We name it separately because, we hypothesise, it needs different design and evaluation criteria (H1, Section~\ref{sec:hyp}).

\subsection{Five-layer stack}
\label{sec:stack}
\begin{figure}[!tbp]
  \centering
  \includegraphics[width=\textwidth]{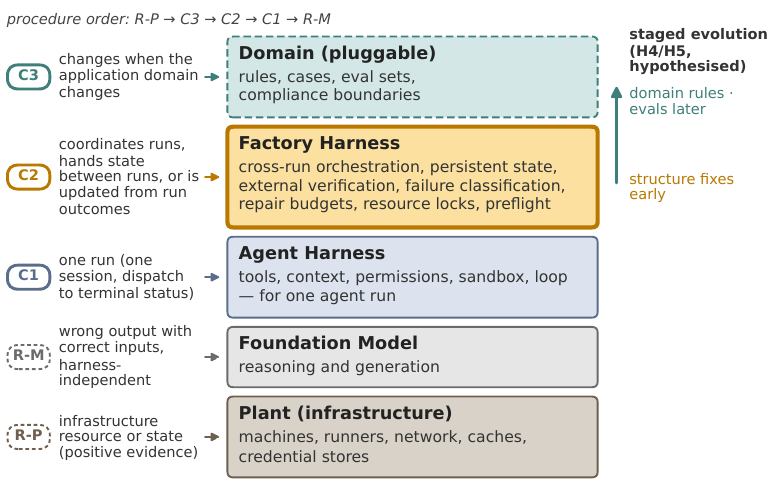}
  \caption{Five-layer stack, with the factory harness highlighted. The markers on the left show where each criterion of manual v1.0 assigns a mechanism. C1 (a single run: one session, from dispatch to terminal status) $\rightarrow$ Agent Harness. C2 (coordinates runs, hands state from one run to a later one, or is updated from run outcomes) $\rightarrow$ Factory Harness. C3 (changes when the application domain changes) $\rightarrow$ Domain. Two residual tests cover the remaining layers: R-P (an infrastructure resource or its state; for incidents, positive evidence of an infrastructure cause) $\rightarrow$ Plant, and R-M (wrong output with correct inputs, independent of the harness) $\rightarrow$ Foundation Model. The tests are applied in the order R-P, C3, C2, C1, R-M (Appendix~\ref{app:codebook}). The right-hand annotation states the staged-evolution proposition, which is a hypothesis tested by H4 and H5. The layering is analytical, not a call or deployment topology; version~1.1 of the manual (Appendix~\ref{app:codebook}) keeps the layers but attributes incidents by a separate procedure.}
  \label{fig:stack}
\end{figure}

Figure~\ref{fig:stack} shows the stack. From top to bottom: the \emph{Domain} layer holds industry rules, cases, evaluation sets and compliance boundaries, and is pluggable; the \emph{Factory Harness} orchestrates runs, keeps state, verifies externally, classifies failures, enforces budgets and locks, and runs preflight; the \emph{Agent Harness} provides tools, context, permissions, sandbox and the loop for one run of one agent; the \emph{Foundation Model} reasons and generates; the \emph{Plant} provides machines, runners, network, caches and credential stores. Two placements differ from related stacks. The domain sits \emph{above} the factory as an input, not inside it as configuration; and infrastructure is a layer of its own, because in E3 a quality result changed with the sandbox, and in the E1 calibration a mismatch in the run environment would otherwise have been read as an agent difference (Sections~\ref{sec:e1} and~\ref{sec:e3}); how common environment causes are, is H6. The stack is an analytical layering for assigning mechanisms and failures, not a call graph or a deployment topology: the plant supports every other layer, and domain content can be executed by the factory or inside a run.

\subsection{Boundary criteria}
\label{sec:criteria}
\paragraph{Unit.} A \emph{run} is one agent-harness session, from the dispatch of a task to the session's terminal status (success, failure, cancellation or unknown). Sub-agents spawned and joined inside that session belong to the run, so sub-agent orchestration inside one session is an agent-harness mechanism, consistent with Barbaste et al.\ and Li et al.~\citep{barbaste2026harness,li2026harnesssurvey}.

\paragraph{Criteria (v1.0).} Given a mechanism (a piece of code, a rule, a configuration item or a process step):
\begin{description}[leftmargin=1.2cm,style=nextline]
  \item[C1] If it concerns only \emph{a single run}, it belongs to the \textbf{Agent Harness}.
  \item[C2] It belongs to the \textbf{Factory Harness} if it (i) coordinates two or more runs (scheduling, ordering, locks, capacity, budgets or retries counted across runs), (ii) holds state that one run writes and a later run reads, or (iii) is selected, parameterised or updated by logic that observes the outcomes of other runs. A static artefact that every run merely loads (harness code, tool definitions, a static instruction file, a fixed prompt template) does not satisfy C2 by persistence alone.
  \item[C3] If it \emph{must change when the application domain changes} (an industry, or a functional domain such as payments), it belongs to the \textbf{Domain} layer.
\end{description}
Two residual tests cover the remaining layers. \textbf{R-P} assigns the \textbf{Plant} to infrastructure resources and their state. A policy \emph{over} a resource, such as a cap, lock or clean-up rule enforced by software that organises runs, is factory, not plant. \textbf{R-M} assigns the \textbf{Model} when generated content is wrong although its inputs and tools were correct, independently of the harness.

\paragraph{Procedure.} The tests are applied in the order R-P $\rightarrow$ C3 $\rightarrow$ C2 $\rightarrow$ C1 $\rightarrow$ R-M, stopping at the first that holds; later hits are recorded as secondary layers (Appendix~\ref{app:codebook}, Table~\ref{tab:codebook-procedure}). Testing infrastructure first catches environment failures that look like quality failures (E3). C3 precedes C2 because a domain rule remains a domain rule when the factory enforces it: a requirement that a medical project name a reviewer for SQL configuration changes is a domain rule \emph{executed by} the factory. Tie-break rules for recurring conflicts are listed in Table~\ref{tab:codebook-tiebreak}.

\paragraph{Incidents: origin, control gap and fix.} For a failure, the layer that produced the error, the layer whose control failed to stop it, and the layer that was changed to remedy it are different questions, and they often have different answers. In E3 a self-test result changed when the sandbox changed, no preflight existed to flag the restricted sandbox, and the remedy was a preflight script; where the error first arose is exactly what the records leave open (Section~\ref{sec:e3}). An incident is therefore coded on three fields:
\begin{itemize}
  \item \texttt{DEFECT\_ORIGIN}: the earliest cause whose correction alone would have prevented the failure;
  \item \texttt{CONTROL\_GAP}: the layer whose control should have detected or prevented the error, or \texttt{NONE} if the error was contained;
  \item \texttt{FIX\_LAYER}: the layer whose artefact the remedy changed.
\end{itemize}
A missing gate is recorded as a control gap and does not move the origin. Without this rule, H3 would be circular: every failure that a factory gate \emph{could} have caught would count as a factory failure. For the origin, R-P requires \emph{positive evidence} of an infrastructure cause. Either the same code behaves differently when only the environment changes, or the record names the cause (quota, regional block, network, disk, missing binary). An unmodified baseline that fails the same way in the same environment shows only that the failure is not a new regression. That fact is recorded separately and does not by itself assign the plant.

\paragraph{Relation to de Macedo.} C1 matches the inner loop of de Macedo's condition T1~\citep{demacedo2026harness}. Under C2, the factory harness is neither de Macedo's \emph{orchestrator} (a fixed graph of steps) nor an \emph{evaluation harness} (which judges a run after the fact): the factory's next step depends on the outcome of external verification and on failure classification, so it is an adaptive outer loop, but one whose unit of observation is a run, not a tool call.

\paragraph{Components versus systems.} The criteria classify components and incidents. They are not necessary and sufficient conditions for a whole system to be a factory harness in the sense of Section~\ref{sec:def}: a fixed sequence of two runs satisfies C2(i), but a system containing it need not be verifiable, recoverable or continuously deliverable. Whether a system meets the definition, and whether it achieves those properties, are separate questions from where its components sit.

\paragraph{Known conflicts in v1.0 and version 1.1.} Applying v1.0 to held-out material exposed conflicts that can change a label (Section~\ref{sec:reliability}). The ordered procedure was used both to classify mechanisms and to locate incident origins, although R-M describes an erroneous generation event rather than the scope of a mechanism, so a model error inside one run satisfies C1 before R-M is reached. Tie-break B4 inferred an agent-harness origin from a harness-side remedy, which reintroduces fix location into the origin. Environment-swap evidence under R-P could assign the plant to an environment that the factory had provisioned wrongly, which B3 assigns to the factory. And the earliest-cause rule presumes that one cause exists. Version~1.1 (Appendix~\ref{app:codebook}) therefore separates two procedures. Mechanisms are classified by type tests (resource, model) and then by scope (C3, C2, C1); C3 requires a stated domain requirement, and C2(iii) applies to the controller that observes outcomes, not to the in-run mechanism it tunes. Incidents are attributed by naming the earliest erroneous artefact or event from the evidence and mapping it to a layer; remedy location is never evidence of origin; environment evidence is recorded separately and assigns the plant only when a resource failed or deviated from what it was supposed to provide; and \texttt{MULTIPLE} or \texttt{UNDETERMINED} with a list of contributing causes replaces a forced primary. The results reported in Section~\ref{sec:reliability} use v1.0; v1.1 has not yet been validated on new material.

\subsection{Structure and evolution: a staged feedback loop}
\label{sec:loop}
\begin{figure}[!tbp]
  \centering
  \includegraphics[width=\textwidth]{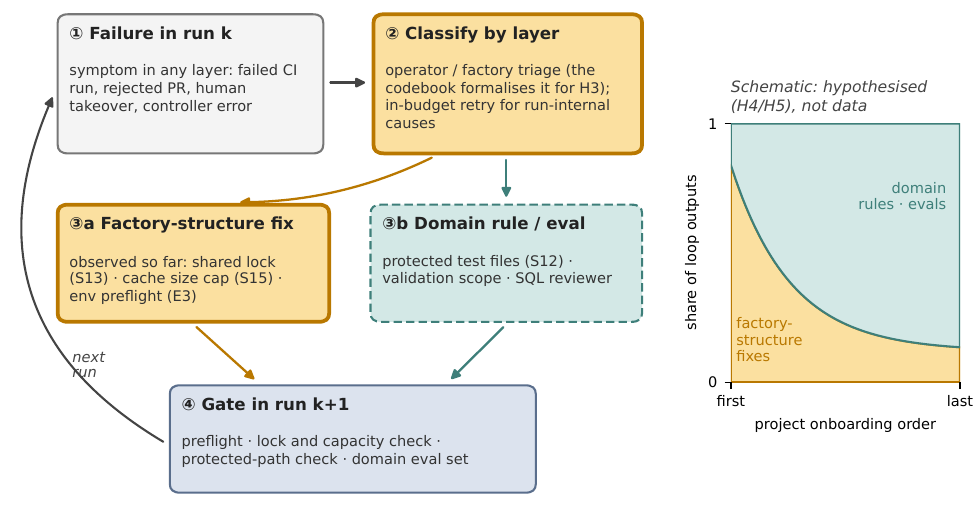}
  \caption{Feedback loop. A failure in run $k$ may show up as a symptom in any layer. It is classified by layer through operator or factory triage; the codebook formalises this step for the H3 analysis and is not a runtime component of System~F. Run-internal causes are retried within the repair budget. Other causes are sedimented either as a factory-structure fix (those observed so far: the shared lock of S13, the cache size cap of S15, the environment preflight of E3) or as a domain rule or evaluation (protected test files of S12, validation scope, SQL configuration reviewer). Either kind then gates run $k{+}1$. Protected test files are an ambiguous case: the mechanism belongs to the Factory and the path list to the Domain (Table~\ref{tab:examples}). The inset is a \emph{schematic} of the hypothesised shift from structure fixes to domain rules over onboarding order, which H4 and H5 test (Table~\ref{tab:hypotheses}). It shows no data, and its curve shape is arbitrary.}
  \label{fig:loop}
\end{figure}

We separate the \emph{structure} of the factory (scheduling, state, the verification interface, the failure-classification frame) from its \emph{rules} (protected files, preflight items, validation scopes). Figure~\ref{fig:loop} shows the loop by which failures become rules: a failure observed in a project is classified by layer, sedimented either as a factory-structure fix or as a domain rule or evaluation, and enforced as a gate in later runs. Feedback loops of this kind are not new: Sengupta et al.\ convert recurring failures into regression tests, template updates, checklist items and compiler rules~\citep{sengupta2026metaeng}, and Factory describes a continuous learning loop~\citep{factory2026factory2}.

Our case does \emph{not} support the strong claim that the loop leaves factory structure untouched. Several lessons became factory mechanisms: two schedulers contending for the same runners led to a shared lock; cache growth led to a size cap; and most of the roughly 40 platform fixes logged as of 2026-10-01 concern factory-layer issues, by the author's own classification made before the codebook existed (Section~\ref{sec:e6}). We therefore state a weaker, staged proposition: \emph{factory structure converges early, after which an increasing share of loop outputs lands in the domain layer}. We have no evidence yet that the structure has converged: at the evidence cut-off the platform repository covered about four weeks (2026-09-06 to 2026-10-03), and fix commits were frequent. H4 and H5 (Section~\ref{sec:hyp}) make the proposition testable; the inset in Figure~\ref{fig:loop} is a schematic of the hypothesis, not data.

\section{Case: \sysF}
\label{sec:case}

\paragraph{Setting.} \sysF{} is a factory-layer pipeline that the author has built and operated since September 2026. It serves six business projects (P1--P6), concentrated in healthcare and elderly care, and uses a self-hosted Git forge as the single source of truth. The platform repository had 202 commits between 2026-09-06 and 2026-10-03, of which 73 have commit messages containing the substring ``fix'' (a rule that also matches words such as ``prefix'') and one begins with ``revert''\ev{E6}. As of 2026-10-01 all six projects ran the same engine version\ev{E6}.

\paragraph{Agent harness layer.} Each project can select one of three managed agent harnesses ($H_A$, $H_B$, $H_C$), each run with its own isolated home directory\ev{E5}. Within $H_A$'s agent family, two execution backends exist: a native protocol backend and a CLI backend\ev{E1}. Engine code does not contain vendor names and external text does not disclose the executor or model\ev{E6}.

\paragraph{Factory harness layer.} The main mechanisms are:
(a) \emph{runner scheduling} (the runner documentation predates the platform repository's first commit): runner tiers are enabled in a fixed order; a burst runner of capacity one accepts only Java CI; a controller runs every 60\,s; tiers share a controller, a state file and a process lock, and the controller fails closed when facts are missing\ev{E6};
(b) \emph{environment preflight}: network-path checks report a \texttt{blocked:network-path-<reason>} status and quarantine the affected path\ev{E6}, and a quality preflight script was added after the sandbox episode in Section~\ref{sec:e3}\ev{E3};
(c) \emph{configuration changes} follow plan $\rightarrow$ apply $\rightarrow$ verify, bound by a summary hash, and exit with code~20 on mismatch; tokens are read only from files with mode 0600; tests run offline with a fake agent\ev{E6};
(d) \emph{trusted-host verification} with signed receipts (Section~\ref{sec:e4});
(e) \emph{run control}: resume, steer and cancel, with process cleanup on parent termination and an \texttt{unknown} status after restart rather than a false success\ev{E6}.

\paragraph{Domain layer.} Domain differences between projects appear as plug-in and configuration fields, for example a validation-scope declaration and a named reviewer for SQL configuration in a medical project, while engine code stays shared\ev{E6}.

\paragraph{Human role.} According to the author's prior talk, the operator reviews only intent and acceptance evidence, while repair budgets, failure classification and back-off are automated\ev{E7}; their parameters are not documented in the E1--E6 sources. In E2 the operator took part in a relay collaboration, so that delivery was \emph{not} unattended (Section~\ref{sec:e2}).

\paragraph{Not yet reported.} Per-project descriptive statistics (runs, tasks, PRs, human interventions, lock and scheduling events) have not yet been extracted from the forge and controller logs; they are the first step of the H3 and H4 extraction.

\section{Preliminary Evidence}
\label{sec:evidence}
This section reports seven bodies of evidence (E1--E7) from \sysF. For each we state what it supports and what it does not. Superscripts such as \ev{E1} key each number to the data appendix (Appendix~\ref{app:data}). None of the comparisons was blinded, and none supports a significance claim.

\subsection{E1: Same-condition comparison of two execution backends}
\label{sec:e1}
\paragraph{Design.} Two execution backends of one coding-agent family (a native protocol backend and a CLI backend) were run with the same model at high reasoning effort, the same prompt, output schema, code tree, permissions and budget. Twelve frozen cases on a small real module, with injected or hand-constructed defects, cover analysis (An1--An3), planning (Pl1--Pl3), defects (Df1--Df3) and constraints (Cs1--Cs3). Each case ran twice per side in fresh sessions with alternating order, giving 48 valid runs (24 per side)\ev{E1}.

\begin{figure}[!tbp]
  \centering
  \includegraphics[width=\textwidth]{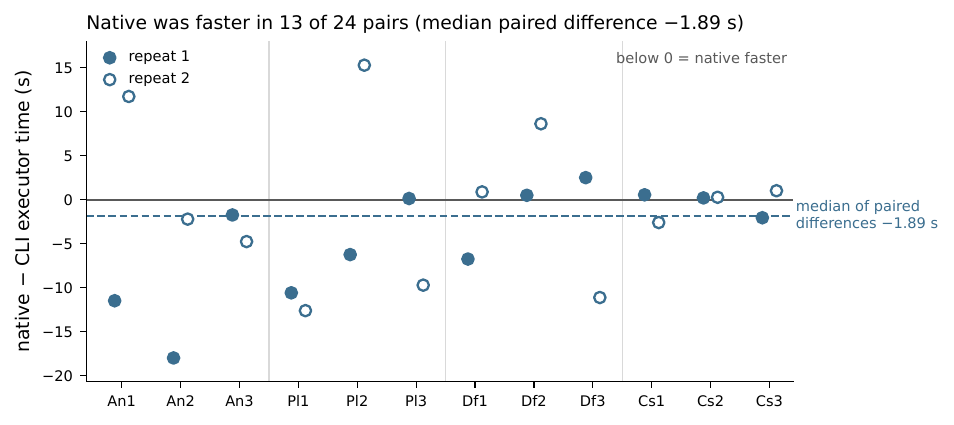}
  \caption{E1 paired executor-time differences (native $-$ CLI) for the 12 cases, two repeats each (filled: repeat 1; open: repeat 2); $n=24$ pairs, same model, prompts, code tree, permissions and budget. Negative values mean the native backend was faster. The dashed line is the median of the paired differences ($-1.89$\,s). Executor time includes runtime and tool execution and excludes external verification. Descriptive only; no significance test.}
  \label{fig:e1pairs}
\end{figure}

\paragraph{Results.} Table~\ref{tab:e1} summarises the results. Analysis, planning and infrastructure-failure cases matched exactly on both sides; independent functional acceptance of write tasks was 10/10 on both sides; structured results were 24/24 on both sides. Strict-rule compliance was 23/24 (native) versus 22/24 (CLI); all three failures (two CLI, one native) concern test placement in case Cs1, where new tests were inserted among existing ones and so failed the grader's byte-prefix rule, while the old assertions were preserved. We do not read the difference as a quality difference. Median executor time (runtime and tool execution, excluding external verification) was 40.96\,s (native) versus 45.28\,s (CLI), about 9.5\% lower for native; P95 (nearest rank) was 60.90\,s versus 61.49\,s. In the paired comparison the native backend was faster in 13 of 24 pairs, with a paired median difference of 1.89\,s in favour of native\ev{E1}. Figure~\ref{fig:e1pairs} shows the 24 paired differences, which range from about $-18$\,s to $+15$\,s. Caches were not cleared and the cases share source code, so we make no claim about speed. A continuity group (Df3 and Cs3, one run per case per side, four runs in total, recorded separately from the 48) completed an analyse $\rightarrow$ external feedback $\rightarrow$ continue-repair sequence in the same session in 82--97\,s (native 82.18 and 87.60\,s, CLI 96.58 and 95.06\,s); this group has no control arm\ev{E1}.

\begin{table}[!htbp]
\centering
\caption{E1: controlled comparison of two execution backends of one coding-agent family (native protocol backend vs.\ CLI backend). Same model (high reasoning effort), prompts, schema, code tree, permissions and budget; 12 frozen cases $\times$ 2 runs per side $=$ 24 runs per side, 48 valid runs in total. Values recomputed from \texttt{results.csv} and checked against \texttt{results-summary.json} and \texttt{REPORT.md}; descriptive only.}
\label{tab:e1}
\small
\begin{tabular}{@{}p{0.46\linewidth}rr@{}}
\toprule
Outcome & Native & CLI \\
\midrule
\multicolumn{3}{@{}l}{\emph{Functional outcomes (runs)}} \\
Analysis (An1--An3) & 6/6 & 6/6 \\
Planning (Pl1--Pl3) & 6/6 & 6/6 \\
Infrastructure-failure runs\textsuperscript{a} & 2/2 & 2/2 \\
Write tasks, independent functional acceptance & 10/10 & 10/10 \\
Structured (schema-valid) result & 24/24 & 24/24 \\
Strict rule compliance\textsuperscript{b} & 23/24 & 22/24 \\
\midrule
\multicolumn{3}{@{}l}{\emph{Executor time (s)}} \\
Median & 40.96 & 45.28 \\
P95\textsuperscript{c} & 60.90 & 61.49 \\
Pairs in which native was faster & \multicolumn{2}{c}{13 of 24} \\
Paired median difference & \multicolumn{2}{c}{native faster by 1.89\,s\textsuperscript{e}} \\
\midrule
\multicolumn{3}{@{}l}{\emph{Continuity group (Df3, Cs3; 1 run per case per side; no control arm)}} \\
Completed analyse $\rightarrow$ external feedback $\rightarrow$ repair & 2/2 & 2/2 \\
Duration range (all 4 runs) & \multicolumn{2}{c}{82--97\,s} \\
\bottomrule
\end{tabular}

\par\smallskip
\begin{minipage}{\linewidth}\footnotesize
\textsuperscript{a}Category membership as reported; case Cs2 (an installation timeout for which no product change is justified), two runs per side; the 10 write runs per side are Df1--Df3, Cs1 and Cs3.
\textsuperscript{b}All three failures concern test placement in case Cs1; the old assertions were retained. The report states this difference cannot be read as higher quality.
\textsuperscript{c}Nearest-rank percentile over 24 values per side, without interpolation.
\textsuperscript{e}Native was also lower on the overall median (about 9.5\%); per-pair differences are shown in Fig.~\ref{fig:e1pairs}. Caches were not cleared; no significance is claimed.
Calibration: an initial trial batch with inconsistent HOME settings was excluded; after a Git porcelain parsing bug, 40 single-round results were re-graded and the 8 repairable runs re-run; the adjudicator accepted all 5 known-correct patches and rejected all unfixed baselines; the final-workspace audit of all 48 sessions found the baseline unmoved and changes confined to the 2 allowed files.
Limitations: token counts are not comparable across backends (last request vs.\ whole turn); review was not blinded; no ticket/PR/CI/merge/deploy stages and no 24\,h stability observation; this is a backend switch within one agent family, not a comparison of different agent harnesses.
\end{minipage}
\end{table}

\paragraph{Calibration and audit.} Several steps preceded these numbers (the adjudicator is described in Appendix~\ref{app:adjudicate}). An initial trial batch, in which the CLI backend inherited the user's home directory while the native backend used a temporary one, was excluded in full, and the formal batch was frozen after the home directory was unified. A defect in the adjudicator (missing \texttt{strip()} on Git porcelain output) caused misjudgements; the 40 single-round results were re-graded with the corrected grader and the 8 runs eligible for a repair call were re-run in fresh workspaces, giving the 48 valid results. In the benchmark's case definitions only Df3 and Cs3 have a repair budget of two calls, and the adjudication script replaces exactly the cases whose budget exceeds one, so the 8 are those two cases on both sides and in both repeats; the re-runs were symmetric across backends. The case definitions and the script logic are in the ancillary material; the re-run manifest itself was not re-inspected. The adjudicator confirmed that prompt, model and code tree were identical across sides; all five known-correct patches passed and all unrepaired baselines were rejected. An audit of the final workspaces of all 48 sessions found that baselines had not moved and changes touched only the two permitted files\ev{E1}. We report these steps because two of them, the home-directory mismatch and the porcelain defect, are themselves factory- or plant-layer failures that would otherwise have been read as agent differences.

\paragraph{Supports.} In this setting, switching the execution backend produced no observed functional difference. The sample bounds what this can show. With 10/10 and 24/24 successes per side, exact (Clopper--Pearson) 95\% lower bounds on each side's success rate are 0.692 and 0.858. These bounds are illustrative: they treat runs as independent trials, whereas the runs are repeats of 5 and 12 cases. They are also bounds for one backend, not an interval for the difference between backends, which we did not estimate; the point is only that samples of this size leave room for success rates well below 1. \textbf{Does not support:} equivalence of different agent harnesses (this is a backend switch within one family), any claim about speed or cost (token accounting differs: the native side records the last request, the CLI side the whole turn), or behaviour on tickets, PRs, CI, merge or deployment. There was no 24-hour stability observation\ev{E1}.

\subsection{E2: A real medical-form defect}
\label{sec:e2}
On a real defect in a medical form, the baseline passed its old tests (10/10), the delivered version passed the new tests (12/12), and the baseline failed 2 of the 12 new tests on Unicode boundaries, so the new tests catch the old defect\ev{E2}. The regression suite at the time comprised 508 Python tests, 552 engine tests (550 passed, 2 skipped), 9 Node tests and 16 front-end tests\ev{E2}. \textbf{Does not support} unattended delivery: the operator took part through relay collaboration, the independent paired run was recorded as not executed, and $N=1$\ev{E2}.

\subsection{E3: \sysF{} versus a general agent desktop application on a cross-layer defect}
\label{sec:e3}
\paragraph{Design.} Using the same model at low reasoning effort, \sysF{} and a general-purpose agent desktop application each implemented one fix for a ``stale clarification'' defect that spans back end and UI. Held-out checks were written before delivery\ev{E3}.

\paragraph{Results.} The original implementation failed 18 of 35 back-end checks and 32 of 38 UI checks (counted by assertion); both fixed versions passed all checks, and the tests added by each side caught the old defect. On the unified host engine, 554 of 556 tests passed and 2 were skipped\ev{E3}. A residual UI difference remained: in one implementation, a page still showed ``clarification needed'' after the question had been answered\ev{E3}. The held-out checks did not cover this state, which illustrates that held-out checks bound quality from below but do not exhaust it. The final check counts were 513 Python, 554 of 556 engine, 11 protocol and 31 front-end tests\ev{E3}. Table~\ref{tab:e3} gives the details.

\paragraph{A result that changed with the sandbox.} \sysF's self-test inside a restricted sandbox reported 9 failures, 10 errors and 2 skips. The same suite in an external sandbox reported 7 failures, 12 errors and 4 skips, the same counts as the baseline (7/12/4)\ev{E3}. What these observations support is limited by what the summary report records. It does not state the environment of the baseline run, and it records counts rather than test-level results, so equal counts do not show that the same tests failed for the same reasons. Nor does it record whether code, dependencies or configuration differed between the restricted and the external run. The change in the result coincides with the change of sandbox and is consistent with an environment contribution, but it does not establish one, and the equality with the baseline at most suggests that the remaining failures are not new regressions. Under the codebook the origin of this incident is therefore undetermined (Table~\ref{tab:examples}, row~3). The preflight was retroactively marked \texttt{BLOCKED} and a quality preflight script was added\ev{E3}. The episode motivated H6 and the evidence rule for R-P (Section~\ref{sec:criteria}), and it shows why the rule is needed: without commit and test-suite identifiers, environment records and test-level results, a plausible environment explanation cannot be told apart from a parallel change.

\begin{table}[!htbp]
\centering
\caption{E3: real cross-layer defect (a stale ``clarification needed'' state spanning back end and UI), implemented once by System F and once by a general-purpose agent desktop application, both with the same model (low reasoning effort). Held-out checks were written before delivery and are counted per assertion. All values author-reported (E3, 2026-09-09 report).}
\label{tab:e3}
\small
\begin{tabular}{@{}>{\raggedright\arraybackslash}p{0.40\linewidth}ccc@{}}
\toprule
Held-out suite & Original & Fix (System F) & Fix (desktop app) \\
\midrule
Back end (35 checks) & 18 of 35 failed & all passed & all passed \\
UI (38 checks) & 32 of 38 failed & all passed & all passed \\
New tests detect the original defect & --- & yes & yes \\
\bottomrule
\end{tabular}

\par\medskip
\begin{tabular}{@{}>{\raggedright\arraybackslash}p{0.55\linewidth}c@{}}
\toprule
Sandbox check (System F fix) & Failed / Errors / Skipped \\
\midrule
Self-test inside restricted sandbox & 9 / 10 / 2 \\
Same check in external sandbox & 7 / 12 / 4 \\
Baseline\textsuperscript{a} & 7 / 12 / 4 \\
\midrule
Unified host engine suite & 554 passed, 2 skipped (of 556) \\
\bottomrule
\end{tabular}

\par\smallskip
\begin{minipage}{\linewidth}\footnotesize
\textsuperscript{a}The external-sandbox counts equal the baseline counts. The summary report attributed the restricted-sandbox failures to environment limits, but it does not state the environment of the baseline run, the suite total, test-level results, or whether code or dependencies differed between the runs, so the attribution is not established (Section~\ref{sec:e3}). The preflight was retroactively marked BLOCKED and a quality-preflight script was added.
Remaining gap: after an answer, one implementation's UI still displayed ``clarification needed'' (detail difference between implementations, as reported).
Final check scale: Python 513, engine 554/556, protocol 11, front end 31.
Limitations: one implementation per side; two real defect classes to date (with E2); not blinded; no resampling.
\end{minipage}
\end{table}

\paragraph{Supports.} On one real cross-layer defect, both implementations met the pre-written held-out checks. \textbf{Does not support} a quality ranking: two problem classes in total (with E2), one run per side, no resampling and no blinding\ev{E3}.

\subsection{E4: Verification outside the agent's sandbox}
\label{sec:e4}
\paragraph{Mechanism.} Verification runs on a trusted host outside the agent's environment and emits a signed receipt (Ed25519, SSH signature format) that binds the repository, the business identifier, the baseline commit, a snapshot digest and an environment digest. Receipts are valid for at most one hour with a 30\,s clock-skew allowance and are checked in order; snapshots are capped at 50{,}000 files or 1\,GB\ev{E4}.

\paragraph{Drills.} Six runs on a real Linux ARM64 host exposed integration problems in sequence: checkout failed without Node; one artefact API version returned empty results and the next was unsupported, so the implementation switched to the contents API; a placeholder environment digest was rejected; a deliberate failure produced \texttt{NOT\_PASSED} and the repaired version \texttt{PASSED}\ev{E4}. The rejection drills are summarised in Table~\ref{tab:e4}: tampered content or a wrong key yields \texttt{SIGNATURE\_INVALID}, an expired receipt is rejected, a changed business identifier, baseline or environment yields \texttt{BINDING}, a changed code tree yields \texttt{STALE\_TREE}, and the original is \texttt{ACCEPTED}\ev{E4}. A second project has been onboarded: 610 engine tests passed with 5 skipped on ARM64 and a bootstrap receipt \texttt{PASSED}; it has not yet been triggered by a real business task\ev{E4}.

\begin{table}[!htbp]
\centering
\caption{E4: rejection matrix of the signed-receipt verifier on a host outside the agent's sandbox, an implemented and drilled instance of a mechanism similar to Proof-or-Stop~\cite{huang2026proofstop}. Receipts are Ed25519 (SSHSIG) signatures binding repository, business ID, baseline commit, snapshot digest and environment digest; validity $\le$1\,h, 30\,s clock skew, ordered checks; snapshot cap 50{,}000 files / 1\,GB. ``Expected'' is the design intent implied by the bound fields; ``observed'' is as reported by the author (E4, 2026-09-09).}
\label{tab:e4}
\small
\begin{tabular}{@{}>{\raggedright\arraybackslash}p{0.38\linewidth}>{\raggedright\arraybackslash}p{0.22\linewidth}>{\raggedright\arraybackslash}p{0.28\linewidth}@{}}
\toprule
Manipulation & Expected & Observed verdict \\
\midrule
\multicolumn{3}{@{}l}{\emph{Rejection drill}} \\
Tampered receipt or wrong signing key & reject (signature) & \texttt{SIGNATURE\_INVALID} \\
Expired receipt & reject (validity) & rejected\textsuperscript{a} \\
Changed business ID, baseline or environment\textsuperscript{b} & reject (binding) & \texttt{BINDING} \\
Changed code tree after signing & reject (freshness) & \texttt{STALE\_TREE} \\
Unmodified original receipt & accept & \texttt{ACCEPTED} \\
\midrule
\multicolumn{3}{@{}l}{\emph{End-to-end drill (real Linux ARM64 host, 6 runs)}} \\
Placeholder environment digest & reject & rejected \\
Deliberately failing change & not passed & \texttt{NOT\_PASSED} \\
Same change after fix & pass & \texttt{PASSED} \\
\bottomrule
\end{tabular}

\par\smallskip
\begin{minipage}{\linewidth}\footnotesize
\textsuperscript{a}Verdict code not recorded in the summary report.
\textsuperscript{b}Whether the three fields were drilled separately, and how many repetitions each drill had, is not specified in the summary report.
Integration issues found during the ARM64 runs: checkout failed without \texttt{node}; one artifact API version returned empty results and the next was unsupported, so the verifier switched to the contents API.
A second project is onboarded (engine tests: 610 passed, 5 skipped on ARM64; bootstrap receipt \texttt{PASSED}) but has not yet been triggered by a real business task.
Known limitations: checks and publication run on the same VM under the same system user; the scheduler uses an administrator token.
The matrix shows mechanism correctness only; without production frequency data it does not show a lower false-pass rate.
\end{minipage}
\end{table}

\paragraph{Relation to prior work.} Proof-or-Stop binds evidence to source-state, policy and command-set hashes, rejects 18 tamper classes offline, and scopes its trust model to a local key on a single host~\citep{huang2026proofstop}; SLSA requires that provenance signing material be inaccessible to user-defined build steps~\citep{slsa2023spec}. E4 is an implemented and drilled instance of a similar mechanism. Its differences are that signing happens on a host separate from the agent's environment, that environment and snapshot digests are mandatory parts of the binding, and that it operates across projects. \textbf{Supports:} correctness of the rejection logic in the tested matrix. \textbf{Does not support} a reduction in false passes: we have no production frequency data. Known limitations: checks and publication run on the same VM under the same system user, and the scheduler uses an administrator token\ev{E4}, so the separation is weaker than SLSA Build L3.

\paragraph{Threat model.} The adversary we have in mind is an agent (or a prompt injected into it) that wants a task accepted without meeting its checks, for example by editing tests, forging a pass, or presenting results computed on a different tree or environment. The agent runs in a sandbox whose file system and credentials do not include the signing key or the verifier's workspace; the verifier recomputes the snapshot and environment digests itself. Under this model, the drills test that forged, stale, rebound or tampered receipts are rejected. They do not test that the agent cannot reach the verifier: no drill had an agent attempt to do so. The model also excludes a compromised scheduler or VM: because checks and publication share a VM and system user, and the scheduler holds an administrator token, an adversary with that access could bypass the verifier.

\paragraph{Open questions.} Two parts of the design are not documented in the sources and remain to be checked; they are open questions, not observed weaknesses. First, the listed receipt fields do not include an identifier of the check set, the verification policy or the verifier version. If these are not bound, a receipt issued under an older policy could be presented after the policy has changed while code and environment stay the same. Second, the verifier executes code and tests that the agent produced. The sources do not describe how those processes are isolated from the signing key, the verifier's workspace and the publication credentials.

\subsection{E5: Substituting agent harnesses}
\label{sec:e5}
All real invocations on 2026-09-20 ran in disposable repositories\ev{E5}. $H_A$ succeeded twice, in about 10\,s and in 13.33\,s, with 26{,}294 and 26{,}524 input tokens, a high context overhead for micro-tasks. In the first round, $H_B$ failed twice with \texttt{resource\_exhausted}, $H_C$ failed twice with HTTP 403 (a regional restriction), and a free-tier model produced one server error after 75.69\,s. After switching models, $H_B$ produced one \texttt{OUTPUT\_INVALID} followed by two correct runs, and $H_C$ produced a correct run (14.9\,s, two tool events)\ev{E5}. Table~\ref{tab:e5} lists the runs.

\begin{table}[!htbp]
\centering
\caption{E5: harness-substitution invocations in disposable repositories (2026-09-20). Three managed agent harnesses (H\textsubscript{A}, H\textsubscript{B}, H\textsubscript{C}), each with a separate HOME. Models differ between rounds, so this is not a same-model comparison. All values author-reported (E5).}
\label{tab:e5}
\small
\begin{tabular}{@{}ll>{\raggedright\arraybackslash}p{0.33\linewidth}>{\raggedright\arraybackslash}p{0.27\linewidth}@{}}
\toprule
Harness & Runs & Outcome & Failure category \\
\midrule
\multicolumn{4}{@{}l}{\emph{First round}} \\
H\textsubscript{A} & 2 & 2 correct ($\approx$10\,s, 13.33\,s; $\approx$26{,}294 and 26{,}524 input tokens) & --- \\
H\textsubscript{B} & 2 & 2 not passed (\texttt{resource\_exhausted}) & quota / resource limit \\
H\textsubscript{C} & 2 & 2 not passed (HTTP 403) & regional access restriction \\
Free-tier model\textsuperscript{a} & 1 & 1 not passed (server-side error, $\approx$75.69\,s) & provider service error \\
\midrule
\multicolumn{4}{@{}l}{\emph{Re-run after model switch}} \\
H\textsubscript{B} & 3 & 1 \texttt{OUTPUT\_INVALID}, then 2 correct & output format \\
H\textsubscript{C} & 1 & 1 correct (14.9\,s, 2 tool events) & --- \\
\bottomrule
\end{tabular}

\par\smallskip
\begin{minipage}{\linewidth}\footnotesize
\textsuperscript{a}The harness used with the free-tier model is not specified in the summary report.
First-round failures are counted as not passed; they do \emph{not} measure harness capability. Failure categories are descriptive labels, not H3 layer codes. We read them as issues of a kind the factory layer has to handle.
The H\textsubscript{A} token counts indicate high context overhead for micro-tasks and are not comparable with E1.
Not covered: real ticket replies, deployment, post-restart acceptance. Each harness was invoked only 1--3 times per round.
\end{minipage}
\end{table}

\textbf{Supports:} each of the three harnesses could be invoked through the same factory interface and produced at least one correct run in disposable repositories; most observed failures (quota, region, output format) are of a kind the factory layer has to handle. \textbf{Does not support} any assessment of the harnesses' capabilities: $N$ is one to three per harness, and real ticket replies, deployment and post-restart acceptance were not completed\ev{E5}.

\subsection{E6: Operations records}
\label{sec:e6}
The runner documentation records two errata: on 2026-09-04 a burst machine was removed, and on 2026-09-24 the 600\,s fallback for tier~2 was found to complete rarely and was replaced by a \texttt{requires\_runners\_busy} condition\ev{E6}. The maintenance log of 2026-10-01 lists about 40 platform fixes traceable by PR number, most of them factory-layer problems: locks, cache-health semantics, ancestry checks, a compare dead loop, a control receipt reported as \texttt{UNKNOWN}, dispatch only after environment readiness and credential cleanup for lost tasks\ev{E6}. Reliability drills measured stop latencies of about 0.327\,s and 0.304\,s; on abrupt parent termination, six processes were cleaned up in about 0.279\,s; after restart, the status was shown as unknown rather than as success\ev{E6}. \textbf{Supports} the descriptive claim that factory-layer problems occurred and were fixed (about 40 in the 2026-10-01 log, classified by the author before the codebook existed). \textbf{Does not support} a claim about their share of all failures; that requires the coding in H3. Business-repository histories have not yet been counted\ev{E6}.

\subsection{E7: Lessons reported by the author}
\label{sec:e7}
A prior talk by the author lists lessons\ev{E7}. They are the author's own earlier account, not evidence independent of E1--E6, and we use them only to motivate hypotheses: agents modified tests or gates, which led to protected files; two schedulers fought over runners, which led to a shared lock and resource checks; a cache grew by a large amount per day (reported by the author as more than 100\,GB per day, not yet checked against monitoring), which led to a size cap; verification must come from somewhere the agent cannot touch; the same model with a different wrapper produced roughly the same quality; and one model writes while another reviews. Agents' tendency to modify failing tests has been measured independently~\citep{zhong2025impossiblebench}. The author's L1--L5 autonomy levels are self-defined and unvalidated.

\section{Developing and Checking the Criteria}
\label{sec:apply}
The criteria were developed on two sets of material (Section~\ref{sec:devset}), frozen as version~1.0 of the coding manual (Appendix~\ref{app:codebook}), and then applied by two independent model coders to units not used in development (Section~\ref{sec:reliability}). Keeping the two apart matters: rules revised on a set of cases cannot also be validated on that set. For the same reason, the revision to version~1.1 prompted by that check is reported only as a sensitivity analysis (Section~\ref{sec:v11}).

\subsection{Development set}
\label{sec:devset}
\paragraph{Documented failures and fixes in \sysF.} Table~\ref{tab:examples} codes 22 items from E1 and E3--E7 on the three incident fields. The items were first coded under version~0 on a single layer field, and that coding informed the rules; after freezing, the author re-coded them with version~1.0 on the three incident fields. Because these items shaped v1.0, the re-coding describes them and does not test the rules. Four rows are excluded, one as study instrumentation and three as mechanisms or domain rules rather than failures. Of the remaining 18 incidents, the defect origin is factory in 10, plant in 5, internal to a run in 2 and undetermined in 1, the E3 episode, whose environment attribution the records do not establish (Section~\ref{sec:e3}). The fix changed a factory artefact in 15. For 10 of the 18 the records do not say which control should have caught the defect earlier. This is itself a finding about the evidence: the platform's own fix records support statements about where errors arose and where they were fixed, but not about which controls failed. The counts reflect selection, because the documented records are mostly operations and platform fixes, so the table is not preliminary evidence for H3. Six incidents are flagged ambiguous. Five are flagged by a named tie-break rule: a per-run value supplied by a cross-run launcher (B3), a provider error that does not name the exhausted resource (B6), agent versus model behaviour (B4), a protection mechanism with a domain-specific path list (B1), and a resource policy that could live in the cache service or the controller (B2). The sixth is the E3 episode, for missing information.

{\scriptsize
\setlength{\tabcolsep}{3.5pt}
\begin{longtable}{@{}>{\raggedright\arraybackslash}p{0.02\textwidth}>{\raggedright\arraybackslash}p{0.30\textwidth}>{\raggedright\arraybackslash}p{0.04\textwidth}>{\raggedright\arraybackslash}p{0.15\textwidth}>{\raggedright\arraybackslash}p{0.14\textwidth}>{\raggedright\arraybackslash}p{0.11\textwidth}>{\raggedright\arraybackslash}p{0.14\textwidth}@{}}
\caption{\textbf{Development set}: failures and fixes documented in E1--E7, re-coded by the author with manual v1.0 on the three incident fields after freezing (Section~\ref{sec:criteria}); an earlier single-field coding under version~0 informed the rules. Rows 2, 8, 9 and 22 are excluded: row 2 is a defect in the study's own instrumentation, and rows 8, 9 and 22 record mechanisms or domain rules rather than failures. Of the 18 remaining incidents, origins are Factory 10, Plant 5, run-internal 2 and undetermined 1; control gaps are \emph{unknown} for 10, none (contained) for 5 and Factory for 3; fixes are Factory 15, Plant 1 and none recorded 2. Six of the 18 are flagged ambiguous. Single coder, items coded from short descriptions; these items were used to develop the rules and are not a test of them. B-codes refer to the tie-break rules in Table~\ref{tab:codebook-tiebreak}.}
\label{tab:examples}\\
\toprule
\# & Item & Src & Defect origin & Control gap & Fix & Ambiguous? \\
\midrule
\endfirsthead
\multicolumn{7}{@{}l}{\textit{(continued)}}\\
\toprule
\# & Item & Src & Defect origin & Control gap & Fix & Ambiguous? \\
\midrule
\endhead
\bottomrule
\endlastfoot
1 & Run batch with inconsistent HOME excluded from the backend comparison & E1 & Factory (B3) & None (caught in trial run) & Factory & \textbf{yes}: AMB-PROVISION \\
2 & Git porcelain \texttt{strip()} defect in the adjudicator; 40 results re-graded, 8 runs re-run & E1 & \multicolumn{3}{l}{\emph{excluded: study instrumentation}} & -- \\
3 & Restricted-sandbox self-test 9F/10E/2S vs.\ external 7F/12E/4S (= baseline); preflight added & E3 & Undetermined (env.\ contribution claimed; baseline env.\ and test-level results not recorded) & Factory (no preflight) & Factory & \textbf{yes}: AMB-INFO \\
4 & \texttt{resource\_exhausted} on $H_B$, first round, 2 runs & E5 & Plant (B6) & None (run not accepted) & Factory (model switch) & \textbf{yes}: AMB-EVIDENCE \\
5 & HTTP 403 regional restriction on $H_C$, 2 runs & E5 & Plant (B6) & None & Factory (model switch) & no \\
6 & Server-side error on a free-tier model, 1 run & E5 & Plant (B6) & None & none recorded & no \\
7 & \texttt{OUTPUT\_INVALID} on $H_B$ after model change (next 2 runs correct) & E5 & Run-internal (B4) & None (schema check caught it) & none recorded & \textbf{yes}: AMB-RUN-INTERNAL \\
8 & Runner tiers share controller, state file and process lock; fail-closed & E6 & \multicolumn{3}{l}{\emph{excluded: mechanism, not a failure}} & -- \\
9 & Network-path preflight with \texttt{blocked:network-path-}\textless{}reason\textgreater{} status & E6 & \multicolumn{3}{l}{\emph{excluded: mechanism, not a failure}} & -- \\
10 & Lock fixes (platform PRs) & E6 & Factory & unknown & Factory & no \\
11 & Cache-health semantics fix & E6 & Factory & unknown & Factory & no \\
12 & Ancestry-check fix & E6 & Factory & unknown & Factory & no \\
13 & Compare dead-loop fix & E6 & Factory & unknown & Factory & no \\
14 & Control receipt reported UNKNOWN after restart (not a false success) & E6 & Factory & unknown & Factory & no \\
15 & Dispatch only after environment is ready & E6 & Factory (ordering) & unknown & Factory & no \\
16 & Credential clean-up for lost tasks & E6 & Factory & unknown & Factory & no \\
17 & Erratum 2026-09-04: one burst machine decommissioned & E6 & Plant & unknown & Plant & no \\
18 & Erratum 2026-09-24: tier-2 600\,s fallback rarely completed $\rightarrow$ \texttt{requires\_runners\_busy} & E6 & Factory & unknown & Factory & no \\
19 & S12: agent edits tests or gates $\rightarrow$ protected files & E7 & Run-internal (B4) & Factory (+Domain list) & Factory (+Domain) & \textbf{yes}: AMB-MECH-CONTENT \\
20 & S13: two schedulers conflict $\rightarrow$ shared lock, resource check first & E7 & Factory & unknown & Factory & no \\
21 & S15: cache growth (author-reported ``100+\,GB/day'', unverified) $\rightarrow$ size cap & E7 & Plant (B2) & Factory (no cap) & Factory & \textbf{yes}: AMB-RESOURCE-POLICY \\
22 & Medical project: SQL configuration reviewer; declared validation scope & E6 & \multicolumn{3}{l}{\emph{excluded: domain rule (mechanism), not a failure}} & -- \\
\end{longtable}
}

\paragraph{Components of three published systems.} Under the version~0 wording, we coded 26 components of HoH~\citep{yan2026hoh}, Sengupta et al.~\citep{sengupta2026metaeng} and Proof-or-Stop~\citep{huang2026proofstop} (Table~\ref{tab:external}, Appendix~\ref{app:external}). Eleven of the 26 were ambiguous (HoH 3 of 7, Sengupta et al.\ 6 of 10, Proof-or-Stop 2 of 9). The commonest problem, in five cases, was a component used within a single run but defined or maintained across runs. Examples are HoH's role invocations and Sengupta et al.'s role layer, agent reviewers, failure arbiter and markdown memory. Other ambiguous cases placed a per-run function in a persistent component, left undefined whether a unit of work was one run or several, mixed a domain-agnostic mechanism with domain content, or depended on plant topology. These cases led to three changes in version~1.0:
\begin{itemize}
  \item the operational definition of a run, with in-run sub-agents counted as part of the run;
  \item the three-part C2 with its exclusion of static, merely loaded artefacts;
  \item ``application domain'' in place of ``industry'' in C3, because functional domains such as authentication were not decided by the old wording.
\end{itemize}
A fourth change came from the incident material: the evidence rule for R-P and the separation of origin, control gap and fix.

\subsection{Held-out reliability check with independent model coders}
\label{sec:reliability}
\paragraph{Design.} After freezing manual v1.0, we assembled 46 units from published material: 26 mechanisms from the source-code study of eleven harnesses and Omnigent~\citep{barbaste2026harness} and from the ETCLOVG survey~\citep{li2026harnesssurvey}, and 20 incidents from Sengupta et al.'s deployment report~\citep{sengupta2026metaeng}, HoH's experiments and multi-day case~\citep{yan2026hoh} and HarnessFix~\citep{chen2026harnessfix}. A separate extraction agent wrote a neutral two-to-five-sentence description of each unit with a page reference, without layer vocabulary. Two coders, each a fresh model instance, then coded all 46 units independently from the descriptions and the manual alone. Neither saw the other's codes, the source papers or this paper. The hold-out is at the level of units, not sources. No held-out unit was used in development, but two incident sources (Sengupta et al.\ and HoH) also contributed components to the development set, so the material is not independent of it. The extraction agent chose units from the five sources as the most clearly described mechanisms and incidents; there was no enumerated candidate frame, so the sample is purposive. The source passage, page, description and both coders' codes for every unit are in the ancillary material. Our protocol allows either two human or two model coders for this check; for the System~F history it requires a human second coder (Appendix~\ref{app:codebook}).

\paragraph{Results.} Table~\ref{tab:reliability} reports agreement. The coders agreed on all 26 mechanism layers (15 agent, 9 factory, 2 plant). On incidents, agreement was 18/20 on defect origin ($\kappa=0.86$), 19/20 on control gap ($\kappa=0.93$) and 19/20 on fix layer ($\kappa=0.91$). The bootstrap intervals are wide and reach down to 0.66. For the mechanism layer the interval is degenerate, because every unit agrees and so does every resample; the Clopper--Pearson interval for the raw agreement of 26/26 is 0.87--1.00 under an assumption of independent units, which the clustering of units in five sources weakens. Each coder flagged 23 of the 46 units as ambiguous, and the flags agreed on 44 units. The reasons separate the sources of difficulty requested by the manual. Missing facts in the description (AMB-INFO) accounted for 10 and 12 of the flags, units bundling several components or events (AMB-GRANULARITY) for 8 and 8, named tie-break rules for 4 and 2, and conflicting or undecided criteria (AMB-CRITERION) for 1 and 1.

\paragraph{Where agreement came from, and where it failed.} Agreement on incidents rests partly on agreeing that the record is insufficient. Both coders coded the origin as undetermined for 8 of the 20 incidents, the control gap as unknown for 9 and the fix as unrecorded or absent for 13; across the three fields, 30 of the 60 field comparisons are of this kind (a descriptive total, not 60 independent observations). Among incidents where both coders gave a substantive code, they agreed on 10 of 11 origins, 10 of 11 control gaps and 6 of 6 fix layers. There were four disagreements. Two stem from missing information, and in one of them (I03) the manual does not define the point of delivery, which is needed to decide whether a required human intervention counts as a contained error. One (I11) concerns whether re-running attempts that failed on infrastructure errors is a factory fix or only containment. One (I16) exposes a genuine conflict between tie-break B4 (a visible harness-side remedy implies an agent-harness origin) and the earliest-cause rule, in a case where the model's omission came first and a harness flaw hid it.

The coders' notes, written independently, also named the same gaps:
\begin{itemize}
  \item whether a persistent session store that a later run \emph{could} read satisfies C2(ii) when no later read is reported;
  \item how to treat delegation to a different vendor's harness inside one session;
  \item bundled environment units combining resources, evaluation content and tools;
  \item the B4 versus earliest-cause conflict;
  \item the undefined delivery point for \texttt{CONTROL\_GAP};
  \item the lack of a \texttt{FIX\_LAYER} code for remedies that change only product code.
\end{itemize}
These become the agenda for version~1.1. In keeping with the freezing rule, they are not applied to the codes reported here.

\paragraph{What this shows and does not show.} The check shows that two coders who were given only the frozen manual and the same descriptions reached the same assignment in most units, and that, in those descriptions, most units the coders flagged were flagged for missing information or bundling rather than for conflicting criteria. Agreement, accuracy and usefulness are different outcomes, and only the first was measured. The coders were procedurally independent (neither saw the other's codes), but their errors need not be independent: they share a model family, the manual and the descriptions. Five limits apply:
\begin{itemize}
  \item The result does not estimate the agreement of human coders. Shared model biases may raise the observed agreement; trained humans may agree more or less.
  \item The descriptions were written by one extraction agent and were not checked for fidelity against the sources. Ambiguity attributed to ``the material'' may come from the sources or from the extraction, and a description can drop exactly the facts (responsibility, call relations, delivery status, environment contract) on which a rule turns.
  \item The mechanism sample contains no domain or model units, so the mechanism $\kappa$ covers only three of the five layers.
  \item Five of the seven HoH incidents (I11--I15) are score patterns or protocol notes rather than narrated failures. Without them, agreement is 13/15 on origin ($\kappa=0.81$), 14/15 on control gap ($\kappa=0.91$) and 15/15 on fix layer (Table~\ref{tab:reliability}b).
  \item The check measures consistency, not utility: whether the codes improve diagnosis or repair decisions is not tested.
\end{itemize}
A replication with human coders on the System~F history, an independent fidelity check of the descriptions, and a utility test comparing repair decisions with and without the three-field coding are the next steps.

\begin{table}[!htbp]
\centering
\footnotesize
\setlength{\tabcolsep}{4pt}
\caption{Agreement between independent model coders on the 46 units (same model family; no coder saw another's codes, the source papers or this paper). Units: 26 mechanisms (Barbaste et al.\ 18, Li et al.\ 8) and 20 incidents (Sengupta et al.\ 8, HoH 7, HarnessFix 5), none used as units in developing v1.0. (a) Main result: frozen v1.0, coders A and B. (b) The same codes without the five HoH units that are score patterns or protocol notes (I11--I15). (c) Sensitivity re-coding under v1.1 by two further fresh coders, C and D. Version~1.1 was written after seeing the v1.0 disagreements on these units, so (c) is not a validation of v1.1. Cohen's $\kappa$ with 95\% percentile bootstrap intervals (5{,}000 unit resamples); units cluster in five sources, so the intervals understate uncertainty about other material. $^{\dagger}$All units agree, so every resample agrees and the bootstrap interval is degenerate; the Clopper--Pearson interval for raw agreement is 0.87--1.00 for 26/26 and 0.78--1.00 for 15/15. ``Both indet.'': both coders coded UNDETERMINED, UNKNOWN, NONE\_RECORDED or NOT\_FIXED. ``Agree, subst.'': agreement among units that both coders coded substantively.}
\label{tab:reliability}
\begin{tabular}{@{}lrrrrr@{}}
\toprule
Field & $n$ & Agree & $\kappa$ [95\% CI] & Both indet. & Agree, subst. \\
\midrule
\multicolumn{6}{@{}l}{\emph{(a) v1.0, coders A and B (main result)}}\\
Mechanism \texttt{layer} & 26 & 26/26 & 1.00$^{\dagger}$ & -- & 26/26 \\
Incident \texttt{DEFECT\_ORIGIN} & 20 & 18/20 & 0.86 [0.66, 1.00] & 8 & 10/11 \\
Incident \texttt{CONTROL\_GAP} & 20 & 19/20 & 0.93 [0.77, 1.00] & 9 & 10/11 \\
Incident \texttt{FIX\_LAYER} & 20 & 19/20 & 0.91 [0.67, 1.00] & 13 & 6/6 \\
Ambiguity flag (all units) & 46 & 44/46 & 0.91 & \multicolumn{2}{l}{flagged: A 23/46, B 23/46} \\
\addlinespace
\multicolumn{6}{@{}l}{\emph{(b) v1.0, without HoH score-pattern and protocol units I11--I15}}\\
Incident \texttt{DEFECT\_ORIGIN} & 15 & 13/15 & 0.81 [0.55, 1.00] & 5 & 8/9 \\
Incident \texttt{CONTROL\_GAP} & 15 & 14/15 & 0.91 [0.71, 1.00] & 6 & 8/9 \\
Incident \texttt{FIX\_LAYER} & 15 & 15/15 & 1.00$^{\dagger}$ & 9 & 6/6 \\
\addlinespace
\multicolumn{6}{@{}l}{\emph{(c) v1.1 sensitivity re-coding, coders C and D (not a validation)}}\\
Mechanism \texttt{layer} & 26 & 25/26 & 0.93 [0.75, 1.00] & -- & 25/26 \\
Incident \texttt{DEFECT\_ORIGIN} & 20 & 17/20 & 0.78 [0.52, 1.00] & 10 & 7/9 \\
Incident \texttt{CONTROL\_GAP} & 20 & 18/20 & 0.85 [0.65, 1.00] & 5 & 13/13 \\
Incident \texttt{FIX\_LAYER} & 20 & 17/20 & 0.79 [0.56, 1.00] & 10 & 7/7 \\
Ambiguity flag (all units) & 46 & 43/46 & 0.86 & \multicolumn{2}{l}{flagged: C 29/46, D 30/46} \\
\bottomrule
\end{tabular}
\end{table}

\subsection{Sensitivity re-coding under version 1.1}
\label{sec:v11}
\paragraph{Design.} Version~1.1 (Section~\ref{sec:criteria}, Appendix~\ref{app:codebook}) was written to resolve the conflicts listed above and frozen (SHA-256 \texttt{87069e50\ldots}). Two further fresh model instances, C and D, coded the same 46 units under v1.1 with the same isolation rules. Because v1.1 was written after seeing the v1.0 disagreements on these units, this is a sensitivity analysis: it shows which labels depend on the revised rules, and it is not evidence that v1.1 is reliable.

\paragraph{Results.} Agreement was lower under v1.1 than under v1.0 (Table~\ref{tab:reliability}c): 25/26 on mechanism layer ($\kappa=0.93$), 17/20 on origin ($\kappa=0.78$), 18/20 on control gap ($\kappa=0.85$) and 17/20 on fix layer ($\kappa=0.79$), and the coders flagged 29 and 30 of the 46 units as ambiguous. Of the 86 field comparisons, 20 changed between the versions:
\begin{itemize}
  \item \emph{Four v1.0 disagreements were resolved in the intended direction.} I16 became a model origin, because remedy location no longer decides the origin; I03 became a contained error under the defined delivery point; I11 became \texttt{NOT\_FIXED}, because re-running is containment; and I08 became undetermined.
  \item \emph{Seven agreed labels changed.} I12 moved from plant to undetermined under the provisioning distinction; I19 moved from agent to run-internal, because a harness-side remedy no longer implies an agent-harness origin; two product-code-only remedies received the new \texttt{PRODUCT} code; and three control gaps moved from unknown to none under the delivery-point rule.
  \item \emph{Nine new disagreements appeared.} They concern whether an organisational response after an incident counts as its fix (I01, I02), whether restoring availability is a fix or containment (I10), whether a unit describes a guard flaw or the earlier error it hid (I17), separating harness from model when the source names only a context limit (I06), aggregate flaw families (I20), missing information on whether regressing changes passed a gate (I09), and whether a transcript store is ``designed to be read'' by a later run (M16).
\end{itemize}
The revisions remove the conflicts they target and make origin codes more conservative, but they move the residual difficulty to the definition of a fix and to the incident's scope. A v1.2 would need a rule for responses that are not tied to one incident and a rule for incident chains. Any version needs validation on units not seen during its revision.

\section{Hypotheses and Evaluation Protocol}
\label{sec:hyp}
The evidence above motivates seven hypotheses. Common data sources are the code forge's issues, pull requests, reviews, CI runs, commits and reverts; \sysF's event, audit and archive exports (read-only transactional exports, mode 0600); and runner-controller state logs. For all coding tasks we will freeze a codebook (Appendix~\ref{app:codebook}) before coding, code independently with two coders (for System~F data, at least one of them human and not involved in operating the system), report Cohen's $\kappa$~\citep{cohen1960kappa} together with the share of units coded substantively, and pre-register the analysis.

In summary:
\begin{description}[leftmargin=0.9cm,style=nextline]
  \item[H1] With the model fixed, switching the agent harness changes functional outcomes less than adding or removing a factory mechanism such as an external verification gate. E1 is consistent with the first half. HoH reports substantial gains when an outer loop is added over fixed harness--model pairs, though with additional compute~\citep{yan2026hoh}; this points the same way but does not test H1, because HoH does not compare harnesses with the model held fixed. The test crosses at least three harnesses with a gate on/off factor and estimates, on a common risk-difference scale, the difference between the gate effect and the largest harness contrast in a mixed-effects model; the decision rests on that difference, and TOST equivalence of the harness contrasts~\citep{schuirmann1987tost} is a secondary result. Case and repetition counts are set in advance by a precision target for that difference, not by a minimum number of executions.
  \item[H2] Verification the agent cannot reach lowers the rate at which defective work is accepted. Because a new gate also makes more failures visible, raw failure counts can rise after it goes live. We therefore apply one common final acceptance to all dispatched tasks, with and without the gate, and hold the content of the checks fixed so that only where they run changes; a gate that also adds checks, compute or permissions is analysed as a combined intervention, and report the true completion rate, the false-acceptance rate, the detection rate, and time and compute cost. The outcome is close to Proof-or-Stop's visible-pass/hidden-fail measure~\citep{huang2026proofstop}. In production, we would estimate it with interrupted time series at each mechanism's go-live~\citep{bernal2017its}.
\item[H3] Most failed deliveries and human interventions in long-running operation have their \emph{defect origin} in the factory or plant layer rather than within a single run. The control gap is analysed separately, so that a missing gate does not count as an origin. The analysis requires a complete incident frame with a pre-registered de-duplication rule, and it keeps undetermined origins in the denominator, reporting the share under both extreme assignments of them. HarnessFix's attribution of agent failures to harness layers is a point of comparison~\citep{chen2026harnessfix}.
\item[H4] The factory-core share of onboarding diffs falls with onboarding order, without a rise in the absolute amount of factory-core change. A falling share alone is not convergence: core changes can double while domain changes grow faster. Because onboarding order is collinear with platform age, the test includes a control for non-onboarding windows; the six existing onboardings are analysed as exploratory and the confirmatory test is reserved for future onboardings.
  \item[H5] After a failure class is sedimented as a rule or evaluation, its recurrence falls. Because a gate can stop defects from escaping without stopping them from arising, we separate defects that arise, defects intercepted by a gate, and defects that escape delivery; fingerprints are linked across renamed paths and gates by a mapping frozen before extraction.
  \item[H6] Without environment preflight, a substantial share of apparent quality failures has its origin in the environment. The share counts only failures with positive evidence of an infrastructure cause, and only when the resource itself failed or deviated from what it was supposed to provide, not when the factory provisioned the wrong environment (v1.1, Part~B2). Non-regressions, where the baseline fails identically, are reported separately and are never counted as environmental on that basis alone. Related categories are flaky tests~\citep{luo2014flaky} and Sengupta et al.'s ``noise'' class~\citep{sengupta2026metaeng}.
\item[H7] (Exploratory.) Higher self-defined autonomy levels go with fewer takeovers per PR and more interceptions by external verification. Levels are assigned from configuration records before outcomes are extracted, and takeover counts are not part of a level's definition, so that the hypothesis is not true by construction.
\end{description}

Table~\ref{tab:hypotheses} (Appendix~\ref{app:protocol}) gives, for each hypothesis, the current support, data source, metric, statistical method and falsification condition.

\section{Threats to Validity}
\label{sec:threats}
\paragraph{Single operator.} The designer, operator and evaluator are the same person, which invites expectation bias; E1 and E3 were not blinded. Mitigations for future work are blinded review, a second coder and pre-registration.
\paragraph{Small samples.} E1 has 48 runs over 12 cases, repeated within cases; E3 is one defect class (two with E2) with one run per side; E5 has one to three runs per harness. We report no significance and do not extrapolate.
\paragraph{Narrow domain.} Six projects concentrated in healthcare and elderly care limit external validity.
\paragraph{Backend versus harness.} E1 compares two backends of one agent family, not different agent harnesses.
\paragraph{Self-defined scale.} The L1--L5 autonomy levels are the author's and are not validated.
\paragraph{Short history.} At the evidence cut-off the platform repository covered about four weeks (2026-09-06 to 2026-10-03); there is no observation longer than 24 hours of continuous operation.
\paragraph{Measurement and recording bias.} Most records were written by the author. Negative results were recorded (\texttt{OUTPUT\_INVALID}, a \texttt{BLOCKED} preflight, errata), but unrecorded failures may exist.
\paragraph{Inconsistent cost accounting.} Token and cost accounting differs between backends, so we compute no savings.
\paragraph{Construct validity.} The boundary between the factory layer and a meta-harness, or the lifecycle layer of Li et al., is not sharp. The held-out check exposed concrete gaps in C2(ii) for persistent session stores that a later run could read, and in cross-vendor delegation within one session (Section~\ref{sec:reliability}).
\paragraph{Model coders and extraction.} The coders in the reliability check are instances of the same model family and read the same descriptions, so their errors may be correlated; their agreement does not estimate human agreement. The descriptions were produced by one extraction agent and were not checked for fidelity against the sources. The units are published descriptions rather than the System~F history, the hold-out is at unit level, the sample is purposive, and the mechanism sample contains no domain or model units. The check measures consistency, not accuracy or usefulness.
\paragraph{Same material for development and validation.} The development set (Tables~\ref{tab:examples} and~\ref{tab:external}) shaped version~1.0 and is not evidence for it. Only the held-out check bears on the frozen rules. The held-out units in turn informed version~1.1, so the v1.1 re-coding (Section~\ref{sec:v11}) is not evidence for v1.1.

\paragraph{Missing descriptives.} Per-project counts of runs, PRs, interventions and scheduling events are not yet reported, so the multi-project setting is described qualitatively.
\paragraph{Provenance of numbers.} The E1 numbers were recomputed from the per-run results file and agree with the benchmark's own summary; an earlier summary of E1 had the two backends' medians, P95 values and strict-rule counts swapped, which this check caught. All other case numbers (E2--E7) are author-reported and keyed to the de-identified appendix; they were not independently re-derived for this manuscript. The per-run E1 data, the statistics script, the held-out units and all four coders' codes are in the ancillary material, so E1 and Table~\ref{tab:reliability} can be recomputed by others; the logs, code changes and environment configurations behind E2--E7 are not released.

\section{Discussion}
\label{sec:disc}
\paragraph{Disagreement with inside-the-harness taxonomies.} Li et al.\ place cross-run persistence, multi-agent orchestration and issue-to-PR pipelines inside the harness~\citep{li2026harnesssurvey}. We do not dispute that usage. Our layering is an analytical distinction: in their terms, the factory harness is the cross-run part of the harness. We name it separately because the two parts seem to call for different evaluations. E1 found no functional difference when the execution backend was switched within one agent family; HoH reports substantial gains when an outer loop is added over fixed harnesses, though with additional compute~\citep{yan2026hoh}. If H1 holds more generally, evaluations of agent harnesses that do not control for the factory layer will misattribute effects, and the distinction is useful; if H1 fails, the case for separate naming weakens.

\paragraph{What the criteria add over existing classifications.} Three existing schemes classify the same material on different axes. ETCLOVG assigns a project or mechanism to the engineering surface it occupies, under a primary-layer rule~\citep{li2026harnesssurvey}. HarnessFix attributes a failure to the harness artefact that shaped the failing step~\citep{chen2026harnessfix}. Sengupta et al.'s arbiter classifies a failure by the corrective action it calls for~\citep{sengupta2026metaeng}. The separation of cause from remedy is not new. HarnessFix traces a failure to the responsible step and its root cause, analyses where the relevant information was produced, propagated, left unexposed, unconstrained or unverified, and maps each diagnosis to a repair operation with an editing scope~\citep{chen2026harnessfix}. Post-mortems in site reliability engineering record root causes, trigger, resolution, detection and action items as separate entries~\citep{beyer2016sre}. Our increment is narrower: three fixed fields (origin, control gap, fix) coded across five layers, combined with an explicit run boundary that separates per-run from cross-run machinery and with an evidence rule for environmental causes. We know of no scheme that combines these three elements, but we have not surveyed the incident-analysis literature systematically. The cross-walk in Table~\ref{tab:codebook-arbiter} makes the difference concrete. The arbiter's ``noise'' class merges infrastructure causes with verifier miscalibration, and its ``bug'' class does not record whether the missing safeguard was a domain test or a cross-run gate. Both distinctions decide which component an operator should change. We do not claim that the criteria improve diagnosis; that is the utility question that remains open (Section~\ref{sec:reliability}). We claim only that they make these distinctions explicit and that two model coders applied them to the descriptions with the agreement reported there.

\paragraph{Relation to HoH, Sengupta et al.\ and Omnigent.} HoH and Sengupta et al.\ are, in our terms, factory harnesses; Section~\ref{sec:apply} classifies their components. HoH is evaluated on three benchmarks and a multi-day greenfield project, with one fixed harness--model configuration per run, and its independent QA is another invocation of the same configuration~\citep{yan2026hoh}. Sengupta et al.'s verification is also largely agent-based, apart from CI, and the authors note that their independence is structural rather than formal~\citep{sengupta2026metaeng}. Neither describes scheduling of shared runners across projects or a separate plant layer. Omnigent addresses composing and controlling several agent harnesses (sessions, policies, sandboxes)~\citep{databricks2026omnigent}; a factory harness is concerned with external verifiability and reliability across runs (receipts, failure classification, repair budgets, resource scheduling). The two can be stacked, and their boundary is the kind of case C2 has to resolve.

\paragraph{Relation to Factory 2.0.} Factory's software factory is a continuous loop from signals to delivery and back~\citep{factory2026factory2}. Our staged proposition is narrower: it predicts \emph{where} loop outputs land over time (first factory structure, later the domain layer), which can be checked against diffs (H4).

\paragraph{When a factory layer is not needed.} If work consists of interactive single sessions with a human reviewing every change, and the sessions share no persistent state, caches or scarce resources, the mechanisms in C2 add little: there is no cross-run state to protect, no shared runner to schedule and the human is the external verifier. Shared repositories, caches or runners can create cross-run dependencies even with a human reviewing every change, and those dependencies are factory concerns. The layer becomes necessary when runs are unattended, numerous, or share resources and projects; the dark-factory level in Shapiro's scale is the limiting case~\citep{shapiro2026fivelevels}.

\section{Conclusion}
\label{sec:concl}
Several systems now place a layer above the agent harness. We argued that the open problem is the boundary: which mechanisms and which failures belong to the agent harness, the factory harness or the domain layer, and what each should be evaluated on. We proposed boundary criteria with an operational definition of a run, together with a three-field coding of failures that keeps defect origin, control gap and fix apart and requires positive evidence for environmental causes. We developed the criteria on our own records and on three published systems, froze them, and found that two model coders of the same family applied them consistently to the same descriptions of 46 held-out units, often by agreeing that a record was insufficient. The check also exposed conflicts in the frozen rules. A revised version that separates mechanism classification from incident attribution removes the conflicts it targets, but in a sensitivity re-coding it gave lower agreement and moved the remaining difficulty to the definition of a fix and to the scope of an incident. Agreement is not accuracy, and neither is usefulness. Whether human coders agree on complete records, whether the coding leads to better repair decisions, and how far a single-operator deployment across six projects generalises remain open. These are the next studies.

\paragraph{Data availability.} The ancillary files of the arXiv submission (\texttt{anc/}) contain the per-run E1 results with case definitions, a script that recomputes every E1 statistic in Section~\ref{sec:e1}, a reference copy of the grader-correction step, the 46 held-out units with their source pages, all four coders' codes, both frozen manuals with their SHA-256 hashes, and the agreement script behind Table~\ref{tab:reliability}. The remaining de-identified appendix (Appendix~\ref{app:data}) is available from the author on request.

\bibliographystyle{unsrtnat}
\bibliography{refs}

\appendix
\section{Data Appendix Index}
\label{app:data}
Two sets of material support the paper. The \emph{ancillary files} (\texttt{anc/}, released with the arXiv submission) are sufficient to recompute E1 and Table~\ref{tab:reliability}:
\begin{description}[leftmargin=0.9cm,style=nextline]
  \item[E1] \path{anc/e1/results.csv} (48 runs, case labels as in the paper), \path{anc/e1/cases.csv} (stage, repair budget and category of each case), \path{anc/e1/e1_stats.py} (recomputes every E1 statistic in Section~\ref{sec:e1}), \path{anc/e1/adjudicate.py} (reference copy of the grader-correction step; it depends on the unreleased runner).
  \item[Coding] \path{anc/reliability/}: the 46 held-out units with source and page, manuals v1.0 and v1.1, the codes of coders A--D, and \path{reliability.py} (agreement, $\kappa$ and bootstrap intervals, with the seed used here). \path{anc/SHA256SUMS} lists the hash of every file.
\end{description}
The \emph{de-identified data appendix}, available from the author on request, contains the remaining sources. Directory names that would identify products or systems have been replaced by neutral names. All numbers in Sections~\ref{sec:case}--\ref{sec:evidence} are keyed to these files by evidence identifier; a per-number traceability table is part of the appendix.
{\raggedright
\begin{description}[leftmargin=0.9cm,style=nextline]
  \item[E1] \path{benchmarks/backend_comparison/}: \path{REPORT.md}, \path{results.csv}, \path{results-summary.json}, \path{continuity-summary.json}, \path{evidence-audit.json}, \path{review-notes.md}, \path{adjudicate.py}, \path{run.py} (report dated 2026-09-07).
  \item[E2] \path{docs/backend/quality-conclusion-2026-09-08.md}.
  \item[E3] \path{docs/backend/real-quality-comparison-2026-09-08.md}, \path{docs/backend/real-quality-comparison-2026-09-09.md}, \path{scripts/quality_preflight.py}.
  \item[E4] \path{docs/backend/trusted-host-verification-2026-09-09.md}.
  \item[E5] \path{docs/multi-harness-plan.md}, \path{docs/multi-harness-e2e-2026-09-20.md}.
  \item[E6] \path{ops/runner-burst/README.md}, \path{docs/maintenance-status.md} (2026-10-01), \path{docs/backend/p0-reliability-2026-09-07.md}, agent instruction file (\path{agent-instructions.md}), platform-repository commit log (202 commits, 2026-09-06 to 2026-10-03).
  \item[E7] De-identified transcript of the author's prior talk.
\end{description}\par}

\section{Layer Coding Manual, Versions 1.0 and 1.1}
\label{app:codebook}

This appendix condenses the coding manual. The full texts are in the ancillary files. \emph{Version history.} Version~0 (unfrozen wording) produced Table~\ref{tab:external} and an earlier single-field coding of the System~F items; both served as development material. Version~1.0 was frozen on 2026-11-03 (SHA-256 \texttt{8438ab6e\ldots}); after freezing, the author re-coded the System~F items with it on the three incident fields (Table~\ref{tab:examples}, descriptive only), and the held-out check in Section~\ref{sec:reliability} was its first application to unseen units. Version~1.1 was written after that check and frozen the same day (SHA-256 \texttt{87069e50\ldots}); its only application so far is the sensitivity re-coding in Section~\ref{sec:v11}. Tables~\ref{tab:codebook-procedure} and~\ref{tab:codebook-tiebreak} give v1.0 as applied in Section~\ref{sec:reliability}; Table~\ref{tab:codebook-v11} lists the v1.1 changes.

\paragraph{Unit of analysis.} A \emph{run} is one agent-harness session, from task dispatch to terminal status. Sub-agents spawned and joined inside the session belong to the run. Two kinds of unit are coded. A \emph{mechanism} unit is a component, rule, policy or process step and receives one \texttt{layer} code. An \emph{incident} unit is a failure, defect, regression or human intervention and receives three separate codes. In the System~F history, incident units are fix commits, failed CI runs (re-runs of the same commit within one hour collapsed), rejected or reverted PRs, and human takeovers, linked by incident identifier.

\paragraph{Three incident fields.} \texttt{DEFECT\_ORIGIN} is where the error first arose: the earliest cause whose correction alone would have prevented the failure. \texttt{CONTROL\_GAP} is the layer whose control should have detected or prevented the error but did not; it is \texttt{NONE} if a control contained the error before delivery. \texttt{FIX\_LAYER} is the layer whose artefact the remedy changed. The fields are coded independently. A coder may not reassign the origin to another layer because a gate was missing; the missing gate is recorded as the control gap. Separating origin from control gap prevents a circular result for H3, in which every failure that a factory gate could have caught would otherwise count as a factory failure.

\begin{table}[!htbp]
\centering
\small
\caption{Criteria and ordered procedure (v1.0, as applied in Section~\ref{sec:reliability}; replaced in v1.1, Table~\ref{tab:codebook-v11}). Mechanisms: apply the steps in order and stop at the first that holds; later hits are recorded as secondary. Incidents: the same criteria locate \texttt{DEFECT\_ORIGIN}, subject to the evidence rule for Plant.}
\label{tab:codebook-procedure}
\begin{tabular}{@{}>{\raggedright\arraybackslash}p{0.10\textwidth}>{\raggedright\arraybackslash}p{0.10\textwidth}>{\raggedright\arraybackslash}p{0.70\textwidth}@{}}
\toprule
Step & Code & Test \\
\midrule
D1 (R-P) & Plant & Mechanism: an infrastructure resource or its state, not logic over it. Incident: \emph{positive evidence} of an infrastructure cause, i.e.\ the same code behaves differently when only the environment changes, or the record names the cause (quota, regional block, network, disk, missing binary, service error). An identical failure of an unmodified baseline in the same environment shows only \emph{non-regression}; it sets \texttt{non\_regression} and does not by itself assign Plant. \\
D2 (C3) & Domain & Content that must change when the application domain (industry or functional domain) changes: requirements, business rules, cases, evaluation sets, compliance boundaries, domain configuration values. \\
D3 (C2) & Factory & At least one of: (i) coordinates two or more runs (scheduling, ordering, locks, capacity, cross-run budgets or retries); (ii) holds state that one run writes and a later run reads; (iii) is selected, parameterised or updated by logic that observes the outcomes of other runs. A static artefact that every run merely loads (harness code, tool definitions, static instruction files, fixed prompt templates) does not satisfy C2 by persistence alone. \\
D4 (C1) & Agent & Confined to a single run, including in-run sub-agents, or a static definition that each run loads. \\
D5 (R-M) & Model & Generated content wrong with correct inputs and tools, independent of the harness. If D4 and D5 cannot be separated: \texttt{RUN\_INTERNAL}. \\
\bottomrule
\end{tabular}
\end{table}

\paragraph{Why this order.} Infrastructure is tested first because environment failures can look like quality failures (E3); the evidence rule keeps that test from absorbing every non-regression. C3 precedes C2 because a domain rule remains a domain rule when the factory enforces it. C1 is the single-run residual of C2, and R-M is the last residual.

\begin{table}[!htbp]
\centering
\small
\caption{Tie-break rules (v1.0). B3, B4, B5 and B7 are revised in v1.1 (Table~\ref{tab:codebook-v11}). A rule that ends in a reason code sets the ambiguity flag.}
\label{tab:codebook-tiebreak}
\begin{tabular}{@{}>{\raggedright\arraybackslash}p{0.05\textwidth}>{\raggedright\arraybackslash}p{0.22\textwidth}>{\raggedright\arraybackslash}p{0.45\textwidth}>{\raggedright\arraybackslash}p{0.18\textwidth}@{}}
\toprule
ID & Situation & Rule & Reason code \\
\midrule
B1 & Mechanism and content change together & Split if separable. Otherwise, mechanism changed $\rightarrow$ by C2/C1; only content changed $\rightarrow$ Domain. & AMB-MECH-CONTENT \\
B2 & Resource vs.\ policy over it & Resource state $\rightarrow$ Plant. Allocation, cap, lock or clean-up across runs $\rightarrow$ Factory. & AMB-RESOURCE-POLICY \\
B3 & Per-run value supplied by cross-run provisioning & Wrongly provisioned $\rightarrow$ Factory. Correct value misused in the run $\rightarrow$ Agent. & AMB-PROVISION \\
B4 & Agent vs.\ model & Harness-side remedy visible $\rightarrow$ Agent. Persists across harnesses with the model fixed $\rightarrow$ Model. Otherwise \texttt{RUN\_INTERNAL}. & AMB-RUN-INTERNAL \\
B5 & Instance per run, definition maintained across runs & Definition updated from run outcomes $\rightarrow$ Factory. Static definition $\rightarrow$ Agent. & AMB-SCOPE \\
B6 & Provider constraint & Origin $\rightarrow$ Plant. Handling is the fix or control. & AMB-EVIDENCE if the resource is unnamed \\
B7 & Multiple causes & Primary is the earliest cause whose correction alone would have prevented the failure. & --- \\
\bottomrule
\end{tabular}
\end{table}

\begin{table}[!htbp]
\centering
\small
\caption{Changes in version~1.1. Each row names a v1.0 problem exposed by the held-out check and the v1.1 rule that replaces it. Version~1.1 adds the fields \texttt{failing\_artefact}, \texttt{contributing}, \texttt{env\_evidence} (SWAP, NAMED, BASELINE\_ONLY, NONE) and, for mechanisms, \texttt{outside\_run\_authority}; \texttt{non\_regression} is subsumed by \texttt{env\_evidence}.}
\label{tab:codebook-v11}
\begin{tabular}{@{}>{\raggedright\arraybackslash}p{0.33\textwidth}>{\raggedright\arraybackslash}p{0.61\textwidth}@{}}
\toprule
v1.0 problem & v1.1 rule \\
\midrule
One ordered procedure for mechanisms and incidents; a model error inside one run satisfies C1 before R-M & Two procedures. Mechanisms: type tests T-P (resource) and T-M (the model itself), then C3, C2, C1. Incidents: name the earliest erroneous artefact or event from the evidence, then map it to a layer. \\
B4 infers an agent-harness origin from a harness-side remedy & Remedy location is never evidence of origin. Model if the inputs were correct or the behaviour persists across harnesses with the model fixed; agent if a harness defect produced the error; otherwise \texttt{RUN\_INTERNAL}. A harness flaw that hid an earlier model error is the control gap. \\
R-P environment-swap evidence vs.\ B3 provisioning & Environment evidence shows contribution, not origin. Plant only if a resource failed or deviated from what it was supposed to provide; Factory if the environment was selected, configured or provisioned wrongly; Domain if the specification omitted a stated requirement; otherwise undetermined (AMB-PROVISION). \\
Earliest-cause rule presumes a single cause & \texttt{MULTIPLE} for independent necessary causes with no earlier one; \texttt{UNDETERMINED} with \texttt{contributing} when the order is unknown. \\
C2(iii) assigns a tuned in-run mechanism to the factory & C2(iii) applies to the controller that observes outcomes; the in-run mechanism it tunes is coded by its own scope, with Factory as secondary. \\
C2(ii) and stores that a later run could read & Satisfied only if a later run reads the state or the store is designed for that; a session-only record is not. \\
C3 and project-specific values & Domain only with a stated business requirement or domain constraint; values that track repository layout, scale or infrastructure go with the mechanism that consumes them. \\
Delivery point undefined for \texttt{CONTROL\_GAP} & Delivery is acceptance by the requester (merge to the protected branch, release, or acceptance by the requesting person); a required human approval step before it is a control. \\
No code for product-only remedies; re-running counted as a fix & \texttt{FIX\_LAYER} adds \texttt{PRODUCT}; re-running, retrying or waiting without changing an artefact is containment (\texttt{NOT\_FIXED}). \\
Sub-agents of another vendor; bundled units & A sub-agent spawned and joined in the session belongs to the run whatever its vendor; a delegate not joined is a separate run. Bundled units are coded by their main component (B8, AMB-GRANULARITY). \\
\bottomrule
\end{tabular}
\end{table}

\paragraph{Components and systems.} Codes classify components and incidents. Whether a whole system is a factory harness in the sense of Section~\ref{sec:def} is a separate judgement, not implied by one Factory-coded component.

\paragraph{Ambiguity and confidence.} Every unit receives an ambiguity flag, a reason and a confidence rating (1 guess, 2 fairly sure, 3 sure). The reason distinguishes missing facts (AMB-INFO), units that bundle several components or events (AMB-GRANULARITY), and criteria that conflict or do not decide (AMB-CRITERION), in addition to the tie-break codes. This split separates ambiguity caused by the material from ambiguity caused by the rules.

\paragraph{Double coding and agreement.} For the System~F history, Coder~A is the author and Coder~B is a person not involved in operating System~F. Model coders, as in Section~\ref{sec:reliability}, are reported separately and never replace Coder~B. Coders calibrate on 20 training units, which are excluded from the agreement sample. Coding is blind to the other coder and to the author's contemporaneous notes. We report Cohen's $\kappa$~\cite{cohen1960kappa} with bootstrap 95\% CIs for each field, per-category agreement and confusion matrices. Post-freeze amendments are versioned and not applied retroactively. HarnessFix~\cite{chen2026harnessfix} and the ETCLOVG survey~\cite{li2026harnesssurvey} attribute failures or projects to harness layers, using LLM-assisted classification with human validation and single-coder multi-label coding respectively, and neither reports inter-rater agreement.

\begin{table}[!htbp]
\centering
\small
\caption{Cross-walk from Sengupta et al.'s four-way failure arbiter~\cite[\S5.3, Table~3]{sengupta2026metaeng} to the incident fields. The arbiter classifies by corrective action and our fields by origin, control gap and fix, so the mapping is many-to-many.}
\label{tab:codebook-arbiter}
\begin{tabular}{@{}>{\raggedright\arraybackslash}p{0.16\textwidth}>{\raggedright\arraybackslash}p{0.24\textwidth}>{\raggedright\arraybackslash}p{0.48\textwidth}@{}}
\toprule
Arbiter class & Most likely codes & Question that decides the alternative \\
\midrule
Bug & origin \texttt{RUN\_INTERNAL}; gap Domain or Factory & Was the violated clause an application rule missing from the eval set (gap Domain) or a cross-run gate (gap Factory)? \\
Spec gap & origin Domain & Did the gap come from the domain-agnostic contract template or compiler (origin Factory)? \\
Noise & origin Plant if evidenced & Is there positive evidence of an infrastructure cause, or a verifier or CI miscalibration (origin Factory)? \\
Contract ambiguity & origin Domain & Does the ambiguity originate in the compiler's rules (origin Factory)? \\
\bottomrule
\end{tabular}
\end{table}

The arbiter has no class for scheduling, lock or state failures. It also merges infrastructure and verifier causes under ``noise'', and these are exactly the causes that H3 and H6 need to separate.

\clearpage
\section{Evaluation Protocol}
\label{app:protocol}
{\scriptsize
\setlength{\tabcolsep}{3pt}
\begin{longtable}{@{}>{\raggedright\arraybackslash}p{0.025\textwidth}>{\raggedright\arraybackslash}p{0.155\textwidth}>{\raggedright\arraybackslash}p{0.11\textwidth}>{\raggedright\arraybackslash}p{0.13\textwidth}>{\raggedright\arraybackslash}p{0.12\textwidth}>{\raggedright\arraybackslash}p{0.17\textwidth}>{\raggedright\arraybackslash}p{0.17\textwidth}@{}}
\caption{Hypotheses and evaluation protocol. ``Current support'' cites author-reported evidence by E-id; \emph{none of it tests the hypothesis}, and apart from E1 the numbers have not been checked against the data appendix. Common data sources are the self-hosted Git forge (issues, PRs, reviews, CI runs, commits, reverts), System~F event, audit and archive exports, and the runner-controller state logs. Unless stated otherwise, every confirmatory analysis is pre-registered before extraction, and every layer assignment uses the frozen codebook (Appendix~\ref{app:codebook}). H4 and H5 state the \emph{staged} form of the feedback loop (Fig.~\ref{fig:loop}). H7 is exploratory. Each decision rule distinguishes support, refutation and an inconclusive result; an interval that is merely wide does not refute a hypothesis.}
\label{tab:hypotheses}\\
\toprule
& Claim & Current support & Data source & Metric & Statistical method & Decision rule \\
\midrule
\endfirsthead
\multicolumn{7}{@{}l}{\textit{(continued)}}\\
\toprule
& Claim & Current support & Data source & Metric & Statistical method & Decision rule \\
\midrule
\endhead
\bottomrule
\endlastfoot
H1 & \textbf{Inner vs.\ outer changes differ in effect pattern.} With the model fixed, switching the agent harness (inner layer) changes outcomes less than adding or removing a factory mechanism such as an external verification gate (outer layer). &
E1: 48 runs, functional acceptance 10/10 vs.\ 10/10, but this was a backend switch \emph{within one agent family}; E3: both implementations passed all held-out checks; E7-S19. HoH's gains from an outer loop~\cite{yan2026hoh} point the same way but do not test H1. &
$\geq$30 closed real defects from PR history, with frozen baselines and held-out tests; design of $\geq$3 harnesses $\times$ gate on/off, same model, $\geq$3 repetitions per cell (at least 540 executions); final sizes set by a precision target for the effect difference; blinded reviewers. &
Held-out pass rate; first-attempt CI pass rate; repair rounds; wall time. &
Mixed-effects model (case random effect) on the risk-difference scale; primary estimand $\Delta$, the absolute gate effect minus the largest absolute harness contrast, with case-bootstrap 95\% CI; secondary: TOST equivalence of harness contrasts with a $\pm$10\,pp margin~\cite{schuirmann1987tost}, McNemar and Cochran's $Q$. & \emph{Support}: CI of $\Delta$ above 0. \emph{Refute}: CI of $\Delta$ below 0. \emph{Inconclusive}: CI contains 0. The $\pm$10\,pp TOST result is reported but does not decide H1. \\
\addlinespace
H2 & \textbf{Verification location.} Verification the agent cannot reach lowers false passes, i.e.\ cases where the agent claims success but held-out or external checks fail. &
E4: rejection matrix (shows the mechanism works, not that it reduces false passes); E7-S12, S17; E1 adjudicator calibration (known-correct patches accepted, unfixed baselines rejected). Proof-or-Stop already measures a related visible-pass/hidden-fail outcome~\cite{huang2026proofstop}. &
PR records where an agent-reported pass is followed by an external CI or receipt failure; commits touching protected paths (matched against the protected list). & A common final acceptance (held-out tests and external CI) is applied to \emph{all} dispatched tasks in both conditions. Report true completion rate (accepted and passing final acceptance / dispatched), false-acceptance rate (accepted but failing final acceptance / accepted), detection rate (failures stopped by the gate / failures), and time and compute cost. Raw failure counts can rise after go-live because detection improves, so they are not the outcome. &
Interrupted time series with segmented regression at each mechanism's go-live~\cite{bernal2017its}; difference-in-differences across projects with and without the mechanism as a secondary analysis (few projects; descriptive). & \emph{Support}: the false-acceptance rate falls (CI of the change below 0) and true completion does not fall by more than 10\,pp. \emph{Refute}: CI of the change in false acceptance at or above 0. \emph{Inconclusive}: CI spans 0, or false acceptance falls while true completion falls by more than 10\,pp (a trade-off, reported as such). A gate that changes check content as well as location is reported as a combined intervention. \\
\addlinespace
H3 & \textbf{Factory/Plant dominance.} Most root causes of delivery failures and human interventions lie in the Factory or Plant layer, not in single-run agent reasoning. &
E6: $\approx$40 platform fixes, most of them factory-layer; E3 sandbox episode; E5 failures were quota, region and format; E1 HOME and porcelain issues. Table~\ref{tab:examples} is illustrative only. &
All failure-bearing units (U1--U4: fix commits, failed CI runs, rejected or reverted PRs, takeovers) in the platform and business repositories, grouped into incidents by a pre-registered de-duplication rule; the incident frame is enumerated before coding. & Share of incidents with \texttt{DEFECT\_ORIGIN} $\in$ \{Factory, Plant\}; distribution of \texttt{CONTROL\_GAP} reported separately; per-layer shares; ambiguity rate; $\kappa$ per field. &
Multinomial proportions with simultaneous 95\% CIs; $\chi^2$ test of layer $\times$ project; Cohen's $\kappa$~\cite{cohen1960kappa} with bootstrap CI. & \emph{Support}: lower 95\% bound of the Factory\,$\cup$\,Plant origin share above 50\% with undetermined origins counted as neither. \emph{Refute}: upper bound below 50\% with undetermined origins counted as Factory\,$\cup$\,Plant. \emph{Inconclusive}: otherwise. \emph{Untestable} if $\kappa$ is below the pre-registered threshold. \\
\addlinespace
H4 & \textbf{Staged structural convergence.} Factory structure converges early, so the factory-core share of onboarding diffs falls with onboarding order. &
E6: 6 projects on one engine version by 2026-10-01; domain differences appear as plugin fields (E6). \emph{Counter-evidence:} S13, S15 and the E6 fixes are changes to factory structure. &
For each project onboarding, diffs in a pre-registered window before and after onboarding, mapped to layers through a path-to-layer map frozen before extraction. &
Factory-core share of changed lines and files per onboarding; absolute factory-core lines and interface changes per onboarding; control: factory-core change rate in equal-length non-onboarding windows (onboarding order is collinear with platform age). &
Spearman $\rho$ between onboarding order and share; negative-binomial regression of factory-core lines with an offset for total lines changed. With $n=6$ onboardings, report effect sizes and CIs; low power is stated in advance. The six existing onboardings have already been seen by the author, so they are analysed as retrospective and exploratory; confirmatory H4 is pre-registered on future onboardings only. & \emph{Support}: $\rho<0$ with CI excluding 0, no comparable decline in control windows, and no rise in absolute factory-core change. \emph{Refute}: CI of $\rho$ excluding negative values, or later onboardings needing a factory-core share at least as large as the first. \emph{Inconclusive}: otherwise (expected with $n=6$). \\
\addlinespace
H5 & \textbf{Sedimentation lowers recurrence.} After a failure class is sedimented into a rule, eval or mechanism, recurrence of that class falls; over time the sediment shifts from Factory to Domain. &
E3: the sandbox episode led to a preflight script; E7: S12, S13 and S15 narratives of failure turned into a mechanism. &
Failure fingerprints (error type, file, gate name) from CI logs and rejection reasons, linked across renamed paths and gates by a frozen mapping, aligned in time with sediment commits (coded \texttt{FIX\_LAYER}). &
Per fingerprint and run (exposure), separately: defects arising, defects intercepted by a gate, defects escaping delivery; Domain share of sediment over time. &
Poisson or negative-binomial model with a post-sediment indicator and a log-runs offset, plus ITS~\cite{bernal2017its}; fingerprints with no sediment serve as controls; regression to the mean is discussed. & \emph{Support}: post/pre rate ratio CI entirely below 1 and larger decline than control fingerprints, stated separately for arising and for escaping defects (a fall in escapes alone supports only the control form). \emph{Refute}: CI entirely at or above 1. \emph{Inconclusive}: CI contains 1. The staged part is refuted if the Domain share of fixes does not rise. \\
\addlinespace
H6 & \textbf{Environment confound.} Without environment preflight, a substantial share of apparent quality failures are environmental. &
E3: counts changed with the sandbox, environment attribution not established; E1: batch with inconsistent HOME excluded. &
Historical failed CI runs; candidate and baseline re-run in a second, recorded environment; commit, suite and environment identifiers and test-level results. & Share of failures with \texttt{DEFECT\_ORIGIN} = Plant under the positive-evidence rule and the provisioning distinction (v1.1); the non-regression share (baseline fails identically) is reported separately and not counted as environmental; both before and after preflight. &
Proportion with Wilson 95\% CI; pre/post comparison by ITS or a difference of two proportions with CI. & \emph{Support}: lower bound of the Plant-origin share above the pre-registered ``substantial'' threshold. \emph{Refute}: upper bound below it. \emph{Inconclusive}: interval contains the threshold. \\
\addlinespace
H7 & \textbf{(Exploratory) Autonomy level.} At higher author-defined levels L1--L5, takeovers per PR fall and external-verification interceptions rise. &
E7-S5 (author-defined scale, not validated). &
Per-PR level, assigned from configuration records before outcomes are extracted, with takeovers excluded from the level definition; takeovers; review edits; interceptions. &
Takeovers per PR; review edits per PR; interceptions per PR. &
Descriptive; Spearman $\rho$ and negative-binomial models with project effects; no confirmatory claim. & None (exploratory); the pattern is described with CIs, and no support or refutation is claimed. \\
\end{longtable}

}

\section{Classification of External Components}
\label{app:external}
Table~\ref{tab:external} gives the component-level classification summarised in Section~\ref{sec:apply}. Page numbers refer to the arXiv PDFs as read on 2026-11-02.

{\scriptsize
\setlength{\tabcolsep}{4pt}
\begin{longtable}{@{}>{\raggedright\arraybackslash}p{0.29\textwidth}>{\raggedright\arraybackslash}p{0.14\textwidth}>{\raggedright\arraybackslash}p{0.10\textwidth}>{\raggedright\arraybackslash}p{0.11\textwidth}>{\raggedright\arraybackslash}p{0.25\textwidth}@{}}
\caption{\textbf{Development set, version~0 wording}: boundary criteria C1--C3 applied to the components of three concurrent systems (single coder, author side; page numbers refer to the arXiv PDFs). The T1--T4 column applies de Macedo's inclusion test~\cite{demacedo2026harness} \emph{in our reading}; $^{\dagger}$ marks harnesses de Macedo classified as yes on all four conditions (Table~4, p.~12). n.a.: the component is state or a record, not a candidate system. 11 of 26 assignments are ambiguous; 7 of the 11 are AMB-SCOPE (a per-run instance whose definition is cross-run).}
\label{tab:external}\\
\toprule
Component (pages) & Criterion fired & Layer & T1--T4 (our reading) & Ambiguous? (reason) \\
\midrule
\endfirsthead
\multicolumn{5}{@{}l}{\textit{(continued)}}\\
\toprule
Component (pages) & Criterion fired & Layer & T1--T4 (our reading) & Ambiguous? (reason) \\
\midrule
\endhead
\bottomrule
\endlastfoot
\multicolumn{5}{@{}l}{\textit{HoH (Yan et al., arXiv:2609.01481)}~\cite{yan2026hoh}} \\
Project Planner /\allowbreak{} Developer /\allowbreak{} QA Tester role invocations (pp. 6--7, 21--22) & C1 (instance); C2 (definition) & Agent /\allowbreak{} Factory & y/\allowbreak{}y/\allowbreak{}y/\allowbreak{}y$^{\dagger}$ (base harness) & \textbf{yes}: AMB-SCOPE: each role is one run of the fixed base harness--model pair, but its prompt template (fixed role instructions + loop-specific context, p. 7) is defined, and held fixed, across loops \\
Fixed Planner$\rightarrow$Developer$\rightarrow$QA order per loop (Algorithm 1) (p. 9) & C2 & Factory & T1 borderline; T4 y & no (our criteria). Under de Macedo the outer loop is neither a fixed-graph orchestrator (the next objective is chosen from $E_{t-1}$) nor a harness (it has no tool loop of its own) \\
Deterministic Runtime: freezes role inputs, enforces permissions, binds evidence, records project state (pp. 5 (Fig. 3), 7, 9) & C2; C1 for per-invocation permissions & Factory & T4 only & \textbf{yes}: AMB-SCOPE: permission enforcement is an Agent-harness function (C1) placed in a component that persists across loops (C2) \\
Artifact state $A_t$ and evidence state $E_t$ carried across loops (pp. 6--8, 9) & C2 & Factory & n.a. & no \\
Versioned project history (workspace commits, issue history, evidence packets) (pp. 3, 15--16) & C2 & Factory & n.a. & no \\
QA candidate frozen read-only; evidence bound to tested candidate (p. 9) & C2 & Factory & T4 y & no; the QA verdict itself is an agent run (row 1) \\
Role-specific tools and skills for the game case (Godot MCP, asset generation, UI/\allowbreak{}UX skills) (pp. 5 (Fig. 3), 15) & C3 (content); C1 (tool interface) & Domain (+Agent) & T2 y (extends base harness) & \textbf{yes}: AMB-MECH-CONTENT: a per-run tool interface (C1/\allowbreak{}T2) whose content is domain-specific (C3) \\
\addlinespace
\multicolumn{5}{@{}l}{\textit{Sengupta et al. (arXiv:2605.25665)}~\cite{sengupta2026metaeng}} \\
Layer 1 Model (pp. 3--4) & none needed (below agent harness) & Model & n.a. & no \\
Layer 2 Role/\allowbreak{}orchestration: assigns roles and execution order (builder, verifier, reviewer, arbiter) (pp. 3--4) & C2 (order, cross-agent); C1 (role prompt per run) & Factory (+Agent) & framework-like & \textbf{yes}: AMB-SCOPE: role prompts are consumed in one run but defined across runs \\
Layer 3 Contract: two-pass compilation, invariants (pp. 3--5 (\S{}4.2)) & C2 (compiler); C3 (contract content) & Factory /\allowbreak{} Domain & n.a. & \textbf{yes}: AMB-MECH-CONTENT: compiler is domain-agnostic; each compiled contract is project/\allowbreak{}domain content \\
Layer 4 Markdown memory (human-approved permanent + rolling sections) (p. 5 (\S{}4.3)) & C2 (persistent); C1 (injected as run context, cf. T3) & Factory & n.a. & \textbf{yes}: AMB-SCOPE: store persists across sessions, but its effect is context selection inside one run \\
Layer 4 Specialization records (payments, booking, authentication, search, \ldots{}) (p. 5 (\S{}4.4)) & C3 & Domain & n.a. & \textbf{yes}: AMB-OTHER: several entries (authentication, search) are functional domains, which the v0 wording (`industry') did not decide; v1.0 uses `application domain' \\
Layer 5 Execution: coding, migration, UI agents (pp. 3--4) & C1 & Agent & y/\allowbreak{}y/\allowbreak{}y/\allowbreak{}y$^{\dagger}$ (underlying harnesses) & no \\
Layer 6 Verification: independence-based (separate builder/\allowbreak{}tester agents), attention-based (role reviewers), CI (pp. 6--7 (\S{}5, Table 2)) & C2 (cross-agent, CI); C1 (each verifier is an agent run) & Factory & T4: y for CI; not for agent reviewers & \textbf{yes}: AMB-SCOPE: verifiers are agent runs organised across agents; only CI is model-independent \\
Four-way failure arbiter (bug /\allowbreak{} spec gap /\allowbreak{} noise /\allowbreak{} contract ambiguity) (p. 7 (\S{}5.3, Table 3)) & C2 (classification routes the next run) & Factory & n.a. & \textbf{yes}: AMB-SCOPE: implemented as an agent role (Layer 2), so one arbiter call is a single run \\
Layer 7 Calibration loop: retros, regression promotion, contract-template updates (pp. 3, 8 (\S{}6)) & C2 & Factory (outputs: Domain or Factory) & n.a. & no \\
Human outer loop: approvals of contracts and permanent memory, unresolved failure classes (pp. 4, 7 (Table 2)) & C2 (intervention point) & Factory & n.a. & no \\
\addlinespace
\multicolumn{5}{@{}l}{\textit{Proof-or-Stop (Huang et al., arXiv:2607.14890)}~\cite{huang2026proofstop}} \\
Agent-output-as-claim; evidence-admission gate for lifecycle transitions (pp. 2--4 (Eqs. 2--3)) & C2 & Factory & T4 y & no \\
Receipts bound to material/\allowbreak{}head/\allowbreak{}story hashes, policy and command-set hashes, receipt identity (pp. 2, 5 (Table 2)) & C2 & Factory & T4 y & no \\
Unattended loop: plan $\rightarrow$ execute $\rightarrow$ review $\rightarrow$ bounded reflection (retry budget) $\rightarrow$ gate (pp. 6, 15, 17) & C1 (plan/\allowbreak{}execute in one story); C2 (budget, gate) & Factory (+Agent) & T1 y; T4 y; T2/\allowbreak{}T3 via host agent & \textbf{yes}: AMB-SCOPE: whether a `story' is one run or many is not defined, so C1's `run' needs an operational definition \\
Independent review verdicts; cross-vendor quorum with explicit degraded single-host fallback (pp. 13--14) & C2 & Factory & n.a. & no \\
Git-native cross-machine handoff and resume with claim boundary (pp. 25--26 (Fig. 7)) & C2 (state); P (machines) & Factory & n.a. & no \\
Local signing key; single-host trust model (excludes compromised runner) (pp. 4, 10) & T2 (key store = Plant; verification policy = Factory) & Plant /\allowbreak{} Factory & n.a. & \textbf{yes}: AMB-RESOURCE-POLICY: the assurance depends on Plant topology (where keys and verifier run), which C1--C3 do not address \\
Offline receipt-bundle adversarial contract: authentic bundle accepted, 18 tamper classes rejected, 10 reason codes reached (p. 10 (Table 5)) & C2 (test of a Factory mechanism) & Factory & n.a. & no \\
Claim-boundary registry /\allowbreak{} prohibited-wording guard (p. 5) & C2 & Factory & n.a. & no \\
Domain packages (future work; schema smoke tests only) (pp. 4, 37) & C3 & Domain (planned) & n.a. & no \\
\end{longtable}

}

\section{Adjudication Script}
\label{app:adjudicate}
The E1 adjudicator (\texttt{adjudicate.py}) checks that prompt, model and code tree are identical across sides; runs functional acceptance on write tasks; checks strict rules (permitted files, test placement, preservation of old assertions); and audits the final workspace (baseline unmoved, changes confined to permitted files). Before the reported results, it was calibrated on five known-correct patches (all accepted) and on unrepaired baselines (all rejected). A defect in its handling of Git porcelain output (missing \texttt{strip()}) was found and fixed, after which the 40 single-round results were re-graded and the 8 runs eligible for a repair call were re-run in fresh workspaces\ev{E1}.

\end{document}